\documentclass[journal]{IEEEtran}
\ifCLASSINFOpdf

\else

\fi

\usepackage{float}
\usepackage{amssymb}
\usepackage{amsmath, amsthm, amsfonts}
\usepackage{amsfonts} % if you want blackboard bold symbols e.g. for real numbers
\usepackage{graphicx} % if you want to include jpeg or pdf pictures
\usepackage{algorithm}
\usepackage{algorithmic}
\usepackage[update,prepend]{epstopdf}
\usepackage{cite}
\usepackage{xcolor}
\usepackage{soul}
\usepackage{caption}
\usepackage{gensymb}
\usepackage{etoolbox}
\usepackage{rotating}
\usepackage{comment}
\usepackage{balance}
\usepackage{booktabs}
\usepackage{multirow}
\usepackage{hyperref}
\usepackage{CJK} 
\usepackage{tikz}
\makeatletter
\patchcmd{\@makecaption}
  {\scshape}
  {}
  {}
  {}
\makeatletter
\patchcmd{\@makecaption}
  {\\}
  {.\ }
  {}
  {}
\makeatother
\begin{document}
\title{Gesture Recognition using Reflected Visible and Infrared Light Wave Signals}
%Gesture Recognition using reflected Light of Visible and infrared LEDs on a photodetector 
\markboth{ARXIV VERSION UPLOADED: JULY 16 2020, YU ET AL.: GESTURE RECOGNITION USING INRARED/VISIBLE LIGHT}{}

\author{Li Yu,~\IEEEmembership{Student~Member,~IEEE}, Hisham Abuella,~\IEEEmembership{Student~Member,~IEEE}, Md Zobaer Islam,~\IEEEmembership{Student~Member,~IEEE}, John F. O'Hara,~\IEEEmembership{Senior Member,~IEEE}, Christopher Crick,~\IEEEmembership{Member,~IEEE}, and Sabit Ekin$^*$,~\IEEEmembership{Member,~IEEE}% <-this % stops a space
%%\thanks{M. Shell was with the Department
%of Electrical and Computer Engineering, Georgia Institute of Technology, Atlanta,
%\thanks{Manuscript received April 19, 2005; revised August 26, 2015.}}%5

		\thanks{Copyright (c) 2020 IEEE. Personal use of this material is permitted. However, permission to use this material for any other purposes must be obtained from the IEEE by sending a request to pubs-permissions@ieee.org. (\textit{*Corresponding author: Sabit Ekin}.)}

\thanks{L.~Yu, H.~Abuella, M. Z. Islam, J.~O'Hara and S.~Ekin  are with the School of Electrical and Computer Engineering, Oklahoma State University, Oklahoma, USA (e-mail: li.yu10, hisham.abuella, zobaer.islam. oharaj, sabit.ekin\{@okstate.edu\}) }

\thanks{C.~Crick is with the Computer Science Department, Oklahoma State University, Oklahoma, USA (e-mail:~chriscrick@cs.okstate.edu).}

}

\maketitle

\begin{abstract}
% As the technological advancements become integral parts of our lives, researchers from multidisciplinary fields have put great attention to develop new sensing methods that is cost-effective, efficient and ubiquitous for human computer interaction (HCI) applications.
% Gesture recognition is considered as an efficient human-computer interface 
% technique. 

In this paper, we demonstrate the ability to recognize hand gestures in a non-contact, wireless fashion using only incoherent light signals reflected from a human subject.  Fundamentally distinguished from radar, lidar and camera-based sensing systems, this sensing modality uses only a low-cost light source (e.g., LED) and sensor (e.g., photodetector). The light-wave-based gesture recognition system identifies different gestures from the variations in light intensity reflected from the subject's hand within a short (20-35~cm) range.  As users perform different gestures, scattered light forms unique, statistically repeatable, time-domain signatures.  These signatures can be learned by repeated sampling to obtain the training model against which unknown gesture signals are tested and categorized.  Performance evaluations have been conducted with eight gestures, five subjects, different distances and lighting conditions, and with visible and infrared light sources.  The results demonstrate the best hand gesture recognition performance of infrared sensing at 20~cm with an average of 96\% accuracy.  The developed gesture recognition system is low-cost, effective and non-contact technology for numerous Human-computer Interaction (HCI) applications.

\end{abstract}

%\IEEEpeerreviewmaketitle

\section{Introduction}

%%%% HCI why Gesture recognition
With the growth of the computer and communication industries, Internet of Things (IoT) and the application of computers in medicine, Human-computer Interaction (HCI) is becoming an increasingly important technological discipline. HCI research is crucial for creating complex, computerized systems that can be operated intuitively and efficiently by people without any formal training. Ideally, it leverages existing and familiar human experiences to make software and devices more comprehensible and usable.
Well-designed HCI interfaces make it convenient to control machines for education, labor, communication, and entertainment environments\cite{IOT1}. Such efforts have gained much attention in recent years. For example, virtual reality allows employees to better understand the nature of their work, especially when it is in an unfamiliar domain. Speech, gesture and handwriting recognition are also highly effective since they leverage common activities of everyday life. As such, they are important topics in applied HCI research \cite{hcin1,hcin2}.
Hand gesture recognition is another natural choice for HCI.  Simple movements of the hand can represent a type of sign language to machines resulting in the execution of complex actions.  As a result, the recognition of hand gestures as a connection between humans and computers is now an active research area \cite{HCI4}. 
\begin{figure*}[!t]
\centering
\includegraphics[width=6.5in]{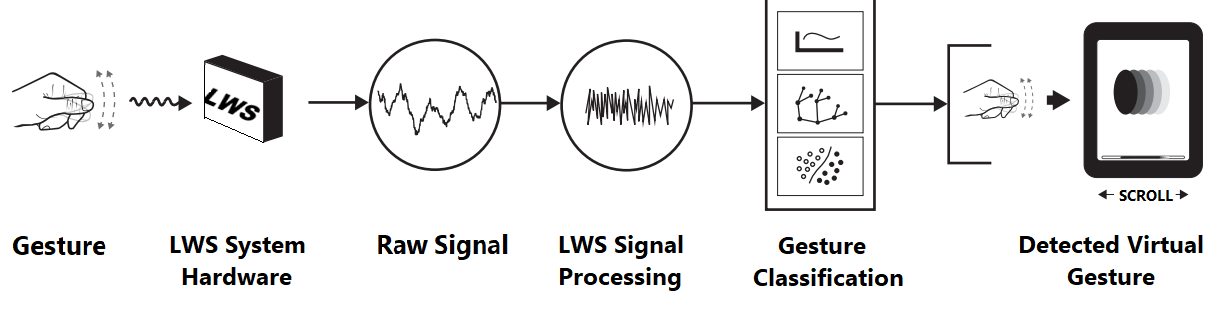}
\caption{Functional overview of light-wave sensing (LWS) based gesture recognition system. }
\label{fig:LS_gesture_system}
\vspace{-3mm}
\end{figure*}

%%%%  Gesture recognition techniques ?
Existing hand gesture recognition techniques can be classified into two groups: wearable sensing and remote (non-contact) sensing. %%%%  wearable sensing ?
In wearable sensing, the user literally wears the sensor(s), which may be installed on a glove or otherwise attached to the hand.   While this sensing mode is both stable and responsive, the sensor(s) must be worn whenever hand movement is to be detected. This inconvenience strips away some of the advantages of wearable sensing. In general, although wearable sensing has higher accuracy, it is simply too inconvenient for many potential users \cite{wear1,wear2,wear3,wear4}. 
%\textcolor{blue}{The wearable sensor relies on the uncontrolled wireless network to transmit data. It is easy to be attacked, because the higher security measurement is hard to applied due to limited bandwidth and processing power. And there are speech disclosure, surreptitious recording and location disclosure issues being recognized in the wearable sensing. \cite{wear5,wear6,wear7}.}
%%%%  remote (non-contact) sensing 

In remote sensing, hand gestures are perceived without any special hardware attached to the hand. The most frequently used sensors utilize radio frequency (RF) waves, cameras, and sound waves. The hand gesture or body motion can be identified by monitoring changes in received signals such as Doppler shifts, signal intensities, or image sequences \cite{light7}. 
% RF
Research on using reflected RF signals (radar) for gesture recognition is relatively mature. In \cite{RF1,RF3,RF4}, authors use the Received Signal Strength (RSS) along with measured phase differences of the received signals as features to identify gestures. In \cite{RF4}, Google presents Soli, the first end-to-end fine gesture recognition and tracking system for HCI using millimeter-wave radar. Soli consists of a system of multiple millimeter-wave radar transmitters and receivers.  The RF-based gesture recognition method is prone to have electromagnetic interference (EMI) and electromagnetic compatibility issues\cite{emirf1,emirf2}.

In imaging-based gesture recognition systems, the input data are images (two- or three-dimensional) and/or videos. The main challenges are separating the objects from the background and feature extraction\cite{image1,image2}.  Deep learning based image classification has attracted much attention with the development of ubiquitous computing power over the last several years. However, deep learning methods usually work with a large amount of training samples and the data needs to be labeled\cite{deep1,deep2}.  Coupled with the large storage and processing requirements of images and videos, this increases the difficulty and complexity of this method. Meanwhile, security and privacy issues also must be taken into consideration \cite{camera5,camera6}.  

Sound-based sensing systems utilize ultrasonic waves and measure the Doppler shift of those waves reflected by the objects.  The velocity of a moving hand, for example, causes characteristic Doppler shifts that serve as a signature to identify activities. The sound-based system is not susceptible to environmental noise and has good accuracy even using an uncomplicated classification algorithm\cite{soundnew1,soundnew2}. Adults cannot hear the ultrasonic frequency, and therefore will not be disturbed. However, the frequencies employed may harm or perturb children and pets\cite{soundnew}.

%%%% light-wave sensing 
The general strategy of using light for sensing has attracted much attention recently due to the advancements in Light Emitting Diodes (LEDs), which now provide unprecedented illumination efficiency and lifetime \cite{light6}. In addition, light can be sensed using simple and inexpensive photodetectors or solar cells. Light signals in general require less processing capability and system complexity, compared to RF systems.  They also suffer far less from cross-technology interference (CTI), owning to the increasing number of RF appliances sharing the same standardized spectrum.  Visible light has already been applied to occupation estimation by analyzing the distribution of reflected and shaded signals\cite{light2}. Objects cast shadows by blocking parts of the light beams from light sources. The shape of a shadow can also be regarded as the pattern in gesture identification \cite{light7,light16}. 
In~\cite{2s}, authors propose a shadow-based hand pose reconstructing system. There are multiple photodiodes placed in the bottom, the binary blockage maps are obtained when the hand gesture blocked the light signal right above the sensors. The hand features are extracted from the blockage maps to build the hand skeleton model to realize the hand gesture recognition and tracking.
Visible light sensing can be applied to detect and identify body and arm gestures based on placing multiple receivers (photodiodes) on the floor or ceiling. However, the interference from obstacles between body and receiver becomes a critical issue in the shadow-based implementation. Therefore, light-wave sensing over shorter distances becomes attractive. Indeed, it appears in many ways that the analysis of signals from reflected light is better suited to application involving small distances \cite{light3,light4}. In~\cite{1s}, authors utilize the infrared light sensing to recognize 6 gestures within 0.5 cm to 7 cm. In~\cite{3s}, authors propose a self-powered gesture recognition system which combines the received visible light signals with setting position. It is a low-cost, highly accurate and stable gesture recognition system within 0.5 cm to 3 cm sensing distance.

%\hl{this paragraph needs help} Indeed, it appears in many ways that the analysis of signals from reflected light is better suited to application involving small distances.  However, reflection-based visible light sensing has \hl{not yet received much attention}. There \hl{is a large quantity} of distance estimation based applications using  infrared proximity sensors \hl{ref 26 is incomplete} \cite{light1,light3,light4}. They analyze the characters, for example, time of arrival or angles of the reflected infrared signals from multiple proximity sensors within a setting to build a skeletal model for localization or gesture recognition. \hl{Move this} In addition, studies conducted by NIST (National Institute of Standards and Technology) indicate that when illuminating towards human skin, the infrared light has larger reflectance value compared to the visible light \cite{nist}. This implies that the reflection signal of infrared light is more sensitive. 
% Our system 

%Considering these existing studies, we developed the idea of utilizing reflected light-wave signal (infrared and visible light) to wirelessly recognize hand gestures through the utilization of  the reflected light signal intensity variation received by photodetector. 

Based on the existing light-based gesture recognition research,  we have developed, for the first time to our knowledge, hand gesture recognition utilizing reflected light-wave (infrared and visible) signals.  Fig.~\ref{fig:LS_gesture_system} depicts our  light-wave sensing (LWS) system.  The main functional components include the LWS hardware, the signal processing algorithms, and the classification algorithms.  LED light sources are used to illuminate the hand. The reflected intensity of the light varies with the movement of the hand and is captured and converted into an electrical current by a commercially-available photodetector.  The time-domain variations of the received (raw) signals are filtered and then classified by using machine learning tools, which are trained by prior captured data sets.  With this modality, we can distinguish different hand gestures with accuracy up to 96\%.  

Our contributions in this study can thus be summarized as follows:

1)	A novel LWS based hand gesture recognition system has been developed.

2)	The gesture recognition system has been implemented in hardware and software sub-assemblies.

3) A comparison between LWS using visible and infrared light has been performed.

4) A system performance summary (classification confusion matrix) has been generated for different distances and environmental lighting conditions.

The remainder of this manuscript is organized as follows. Section~\ref{sec:System_Design} presents the principles and system design, both in terms of hardware and software algorithms utilized.  Section~\ref{sec:Results} presents the  evaluation of the LWS system, and some brief related discussion. Finally, Section~\ref{sec:Conclusion} presents the conclusions and future work.

\section{System Design and Implementation}
\label{sec:System_Design}
\begin{figure}[!t]
\centering
 \includegraphics[width=1\columnwidth,height=6cm ]{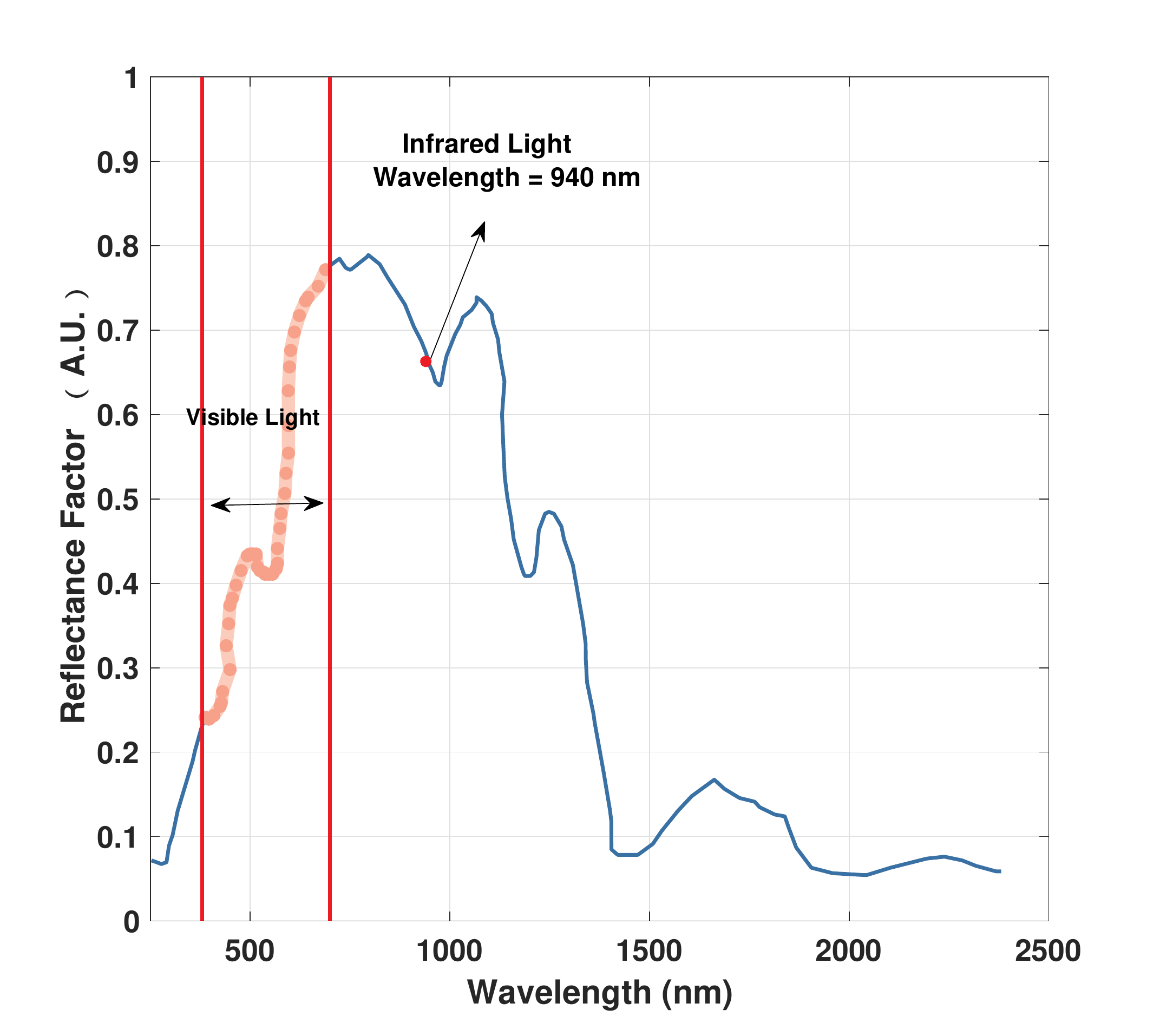}
\caption{Spectral reflectance variability of human skin (regenerated from NIST study \cite{nist}).}
\label{fig:NIST_reflectance}
\end{figure}

\subsection{Influence of the Reflectance Spectrum}

The spectral reflectance of human skin offers unique opportunities for non-contact sensing applications.  Indeed, it can serve as an identifying signature. For example, imaging a human face with hyperspectral cameras provides very broad reflectance spectra, which can be divided into numerous narrow bands.  Each of these can be used to increase the accuracy of face recognition. A NIST (National Institute of Standards and Technology) project collected measurements from 28 human subjects and calculated the spectra for their reflectance measurements over the 250~nm to 2,500~nm wavelength range \cite{nist}. Fig.~\ref{fig:NIST_reflectance} shows the reflectance spectrum for the mean of all the samples. The spectrum exhibits the variations and scales of reflectance factors which are critical for the aforementioned applications.  While hyperspectral information is not required for hand gesture recognition, it does indicate what wavelengths LWS might be implemented at most effectively.  In hand gesture recognition, higher reflectance translates to a larger reflected light intensity for a fixed power level. This feature is beneficial to achieving greater dynamic range in sensing and to reducing power consumption. From the NIST spectrum, the visible/infrared wavelength range of 600~nm to 1,200~nm has the most significant reflectance. Hence, we began LWS using visible and invisible infrared light sources. 

\begin{figure}[!t]
\centering
\includegraphics[width=3.4in]{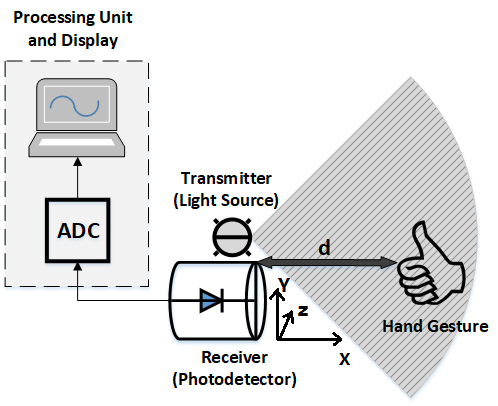}

\caption{Hardware overview of the light-wave sensing based system.}
\label{fig:System_model}
\end{figure}

\subsection{Sensing Hardware}
\begin{figure}[!t]
\centering
\includegraphics[width=3.4in,height=8cm]{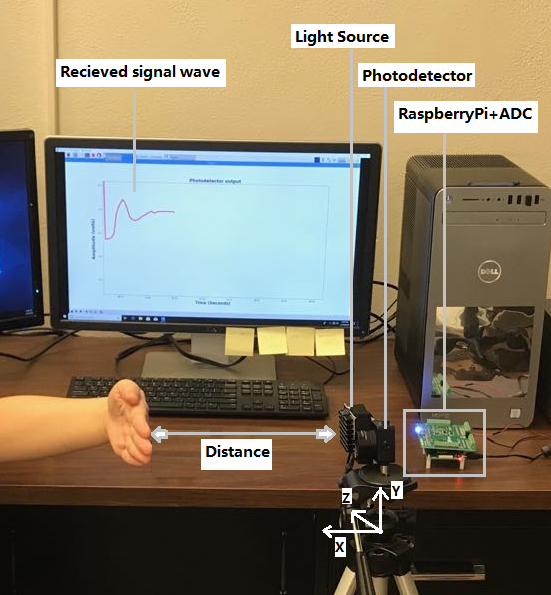}
\caption{{The experimental setup of our light-wave sensing  system.}}
\label{fig:Hardware_setup}
\end{figure}

Our hardware schematic is shown in  Fig.~\ref{fig:System_model} {and the experimental setup is shown in} Fig.~\ref{fig:Hardware_setup}. The LWS hardware consisted of one photodetector as receiver, a digital signal processing (DSP) unit to convert analog data to digital modality, light sources (visible or infrared) as transmitters and finally an electronic unit that could process and store the received digital data.   The light sources consisted of visible and infrared LEDs. The infrared light source was invisible 940 nm IR lamp board with light sensor (48 black LED illuminator array) having 30 ft range and 120\degree~wide angle beam\cite{transmitter}. The visible light source was 25 white 5 mm LED array\cite{vistran}. A Raspberry Pi miniature computer with a PiPlate ADC circuit handled the data collection and digitization. A commercial photodetector served as the detector. The photodetector was Thorlabs PDA100A with spectrum responsivity 340 - 1100 nm, bandwidth 2.4 MHz and area 100 mm$^2$\cite{detector}. The gain of the PDA100A was the same in the visible and infrared measurements (20 dB) in this paper.

\subsection{Measurement Procedure Overview}
In operation, volunteers perform gestures at a distance $d$ in front of the receiver (and transmitter).  As the hand makes different gestures, the photodetector records light intensity waves that are unique according to the changing distance, shape, and scattering cross section of the subject's hand.  The transmitter is pointed such that the hand is centered in the brightest part of the transmitter's beam.  This was visibly obvious in case of the visible light source, whereas an infrared monitor was used to optimize aiming while using the infrared source.  

\begin{figure}[t]
\includegraphics[width=1\columnwidth]{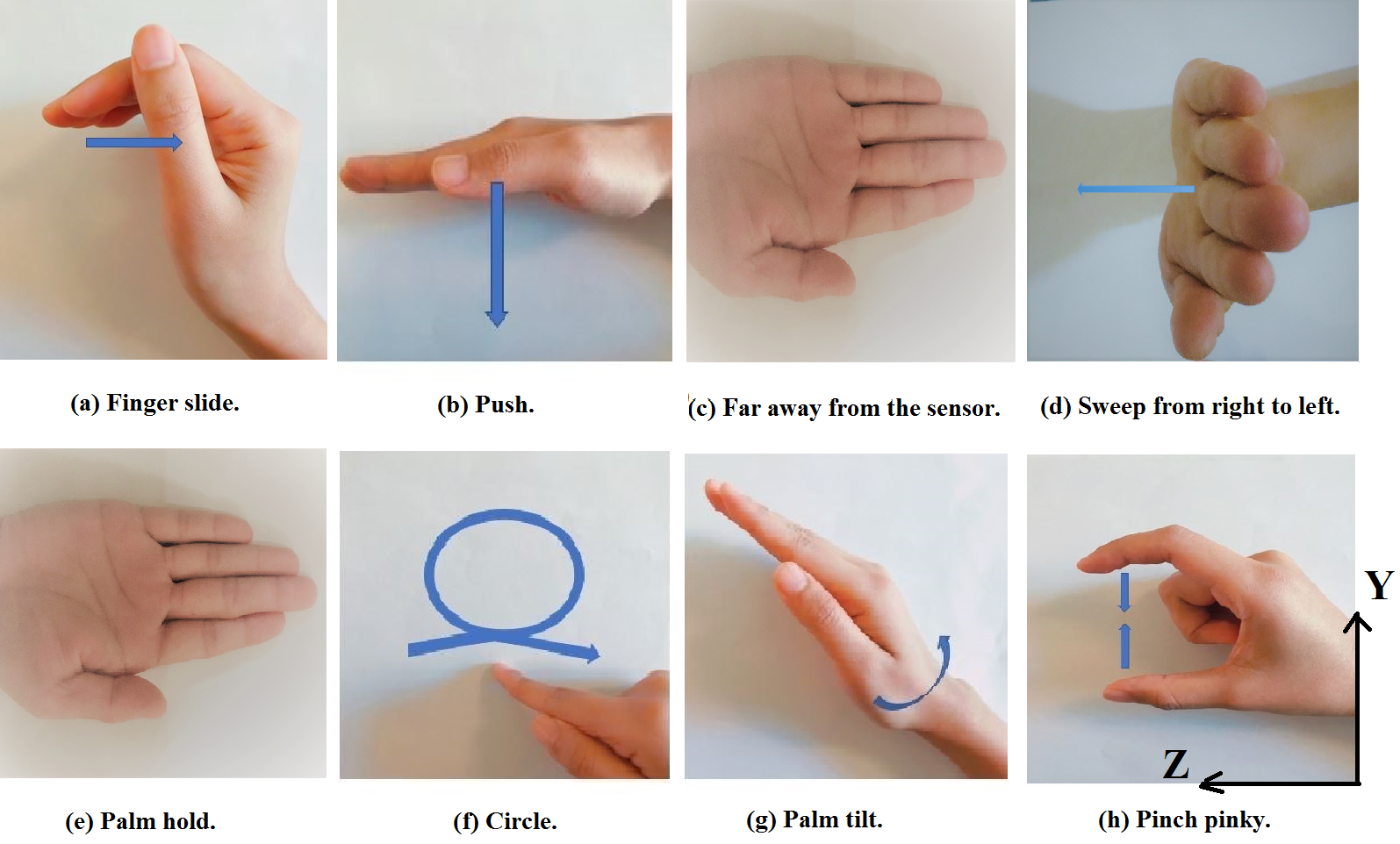} 
\centering
\caption{ {Gesture set.} }
\label{fig:Gesture_set_1}
 \vspace{-3mm}
\end{figure}

Volunteers were asked to perform gestures (Fig.~\ref{fig:Gesture_set_1}) in the designated area in front of the photodetector and light source.  Each gesture could be finished in 2-3 seconds but was recorded for 6 seconds at a sampling rate of 100~Hz, resulting in individual gesture data sets of approximately 600 bytes (single precision). The digital data was then processed and classified offline using the algorithms mentioned in the next subsections. A video that shows the environments, experimental setup and the data collection process was taken (see video link at \url{https://youtu.be/OStciFfvZa0}).

\subsection{Signal Processing}

\begin{figure}[t]
\includegraphics[width=0.32\textwidth]{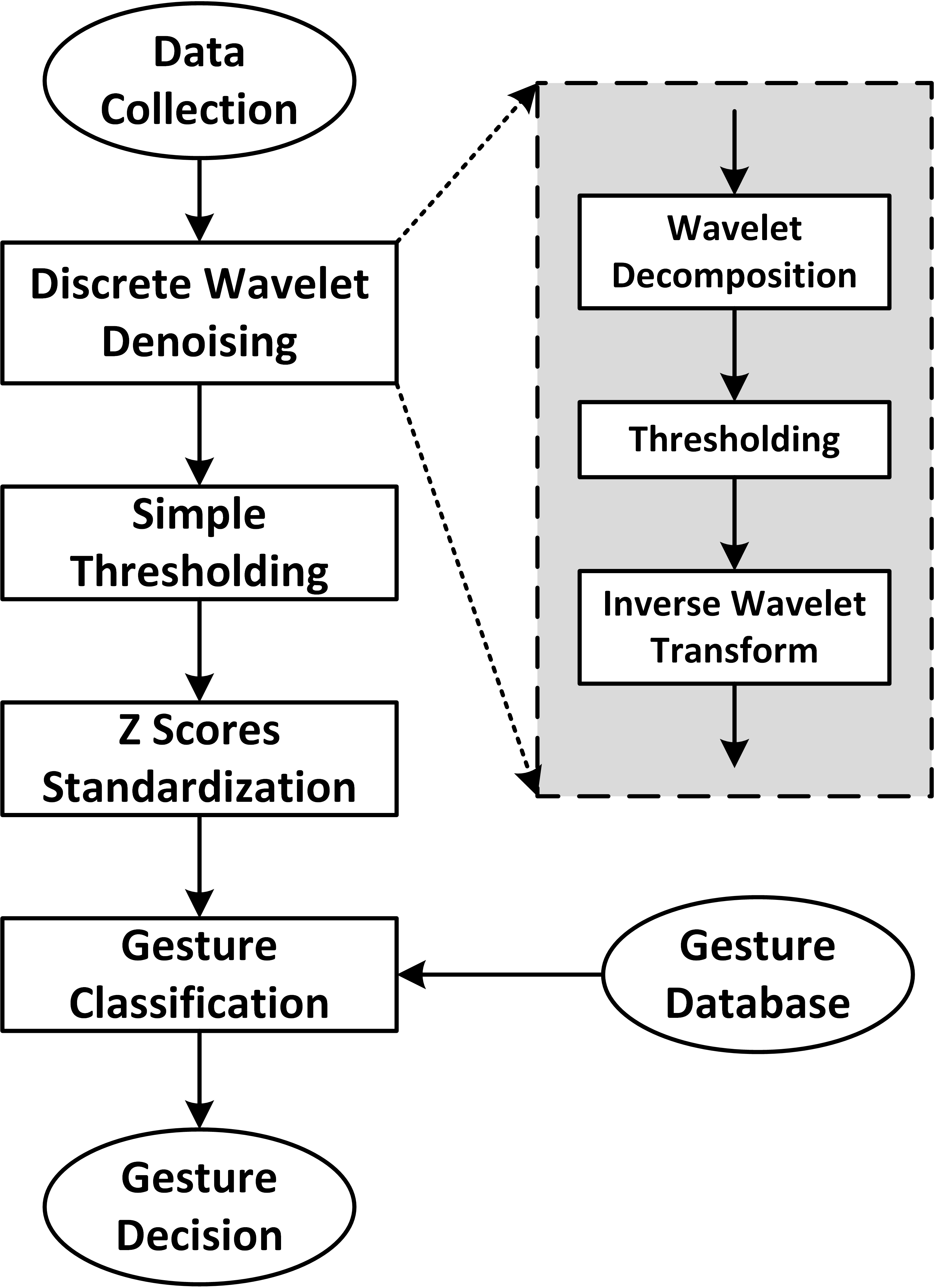} 
\centering
\caption{{Flow diagram of the steps and algorithms used to prepare measured data for gesture classification.}}
% \vspace{-0.6cm}
\label{fig:Flow_Diagram}
\vspace{-4mm}
\end{figure}

In order to extract the patterns and important signal features resulting from different gestures, and to remove redundant information and noise from the received signal, multiple signal processing algorithms are needed. A flow diagram of the sequential operations performed on the raw data is presented in Fig.~\ref{fig:Flow_Diagram}. First, discrete wavelet denoising was used for noise and interference removal.  Then, a simple thresholding scheme was used to segment the long received data stream and mark the beginning and the end of a hand gesture.  Then, Z scores was used to standardize the signal amplitude based on the variance of the data. This normalized the data collected in varying distance and background lighting conditions.  Finally, the data was classified to a certain gesture based on the database of available gestures. Each algorithm is explained in details as follows.

\vspace{2mm}
\noindent\textbf{System Noise and Denoising}

\begin{figure}[t]
\includegraphics[width=0.48\textwidth]{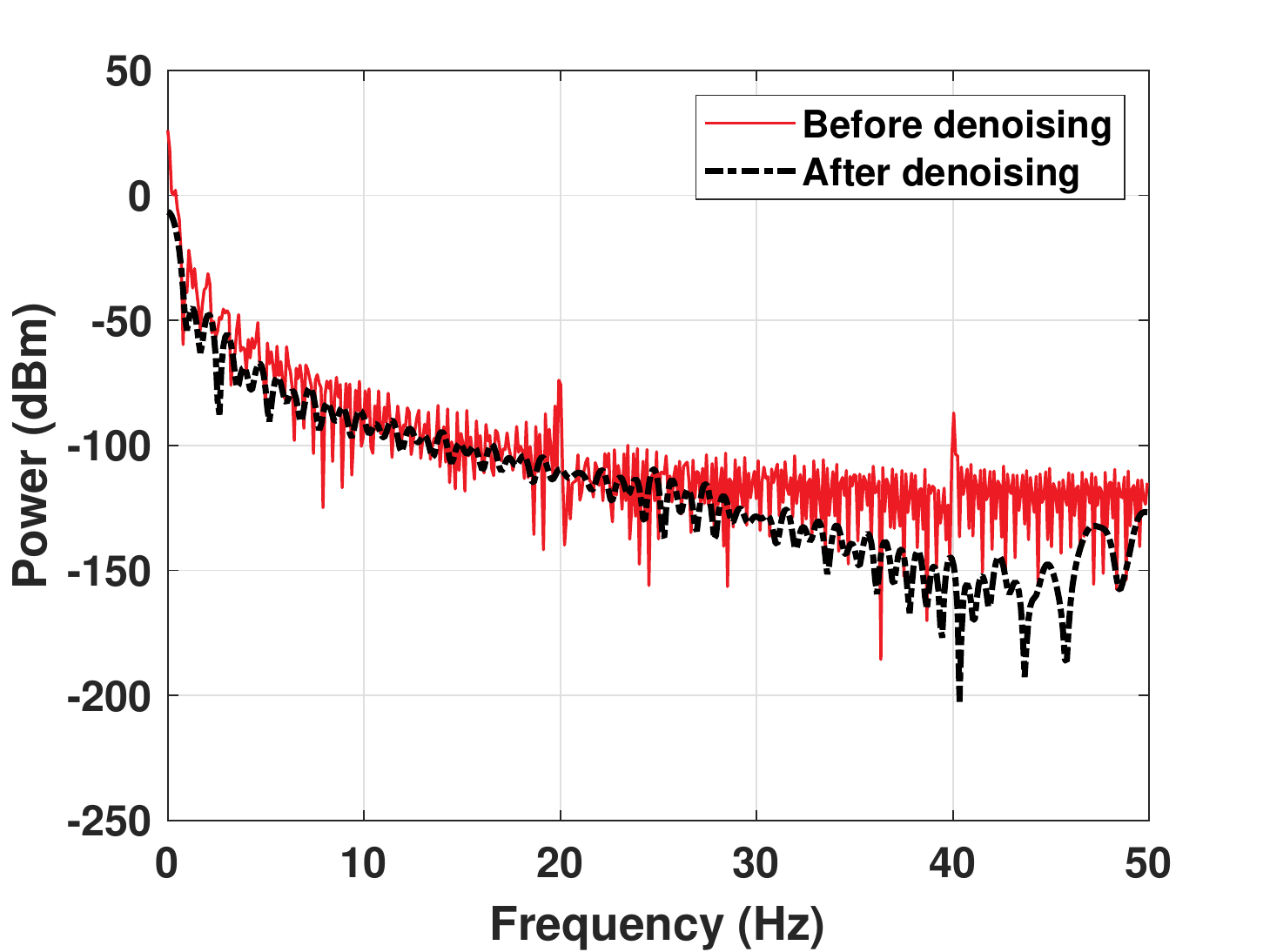}
\centering
\caption{Power spectrum of the received visible light signal at 20~cm} before and after Discrete Wavelet Denoising block.
\label{fig:Denoising_effect}
\end{figure}

The raw signal is usually corrupted by noise that distorts the significant signal features, especially when the amplitude of the reflected signal is relatively low.  Two noise features that were obvious in the frequency domain appear to be due to the flicker of ceiling lights, evident at 120~Hz,  and the flicker of nearby computer monitors, evident at 60~Hz. The magnitude of the Fourier transform of the received signal is shown in Fig.~\ref{fig:Denoising_effect}.  The flicker noise sources appear as peaks at 20 Hz (for the 120 Hz signal) and 40 Hz (for the 60 Hz signal) due to spectral folding caused by the 100~Hz sampling rate being below the Nyquist criterion.

In order to denoise the signal, the Discrete Wavelet Transform (DWT) was used \cite{dwtn1}.  The wavelet thresholding method has been proven remarkably adept in signal denoising in various published research works, including ECG (electrocardiogram) denoising \cite{ecgnew}. The ECG signal is similar to our gesture data in that the heart activities are variable with time on similar scales and indicate different health states based on the various time-domain waves.  The DWT is highly useful in analyzing non-stationary signals since it provides both a time- and frequency-domain representation of the {signal} \cite{dwtnew1}.

\vspace{2mm}
\noindent\textbf{Discrete Wavelet Denoising}

The wavelet thresholding method removed the noise by forcing the DWT coefficients of noise to zero.  Coefficients of noise were distinguished from those of the meaningful part of the signal by their relatively small magnitude. %The coefficients of noise are relatively small compared with the meaningful part of signal, which provides the principal distinction used for filtering. Therefore, the coefficients with small magnitude can be regarded as noise and need to be set to zero. 
Wavelet thresholding means that each DWT coefficient is compared to a threshold to determine whether it is a part of the desired signal or not. Thresholding is usually applied to the DWT detail coefficients which are related to high frequency noise. When the value of a coefficient was found to be less than the threshold, it was forced to zero.  Following this operation, the inverse wavelet transformation used the remaining non-zero coefficients to produce the denoised signal.
Coefficients larger than the threshold were dealt with differently according to the thresholding scheme employed, whether it was `soft' and `hard'.  More details of these two thresholding methods are available in \cite{thold}.
We applied soft thresholding in our method.  We have empirically observed that soft thresholding produces a more mathematically tractable signal, and also one that is easier to interpret visually.  Specifically, the soft method eliminates certain signal `blips' that survive with hard thresholding.

\begin{figure}[t]
\includegraphics[width=0.5\textwidth,height=6cm]{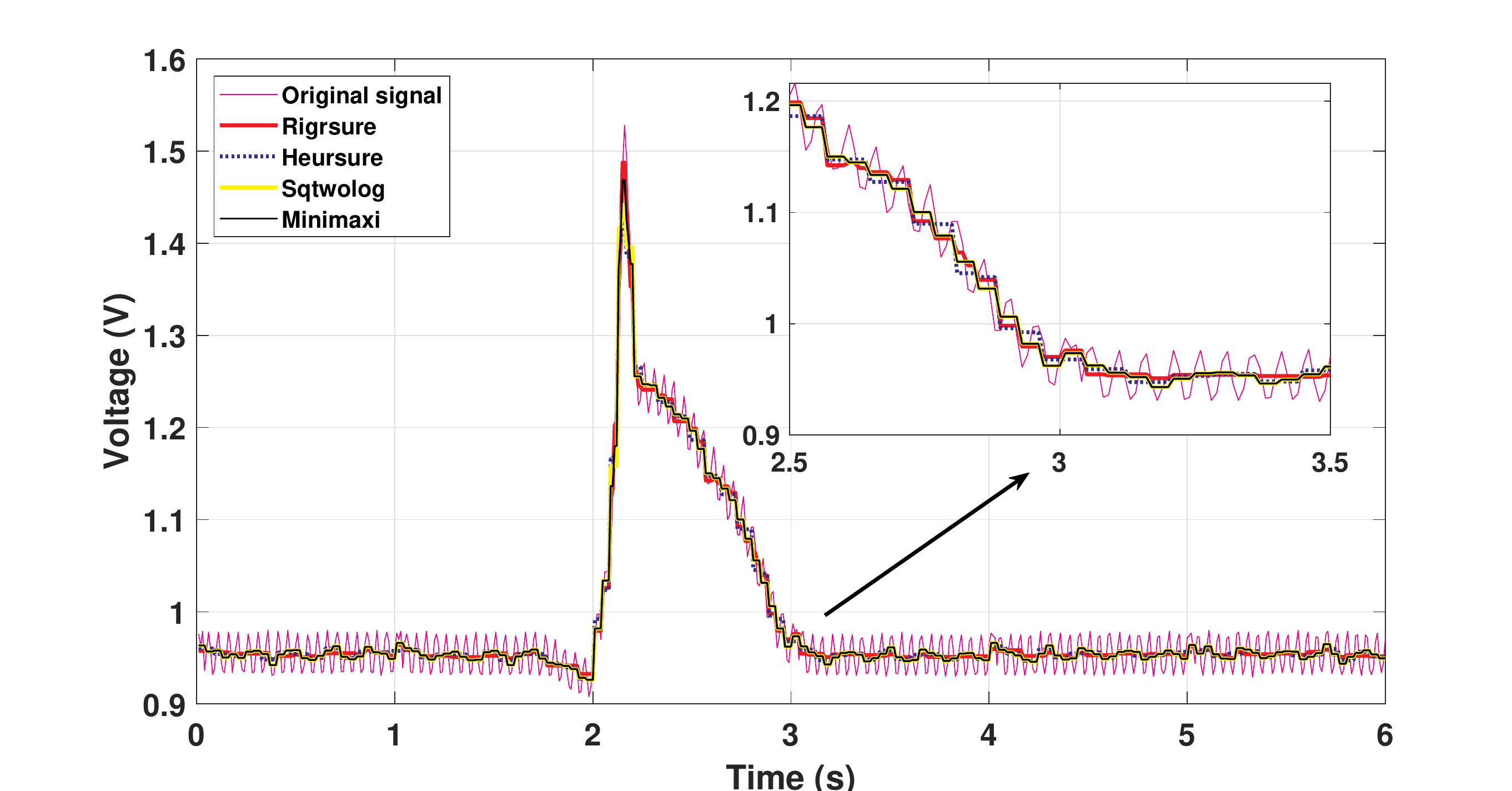}
\centering
\caption{The denoised wave of received visible light signal at 20~cm with four thresholding methods.}
\label{fig:noise_p}
\vspace{-4mm}
\end{figure}

\begin{figure}[t]
\includegraphics[width=0.5\textwidth,height=6cm]{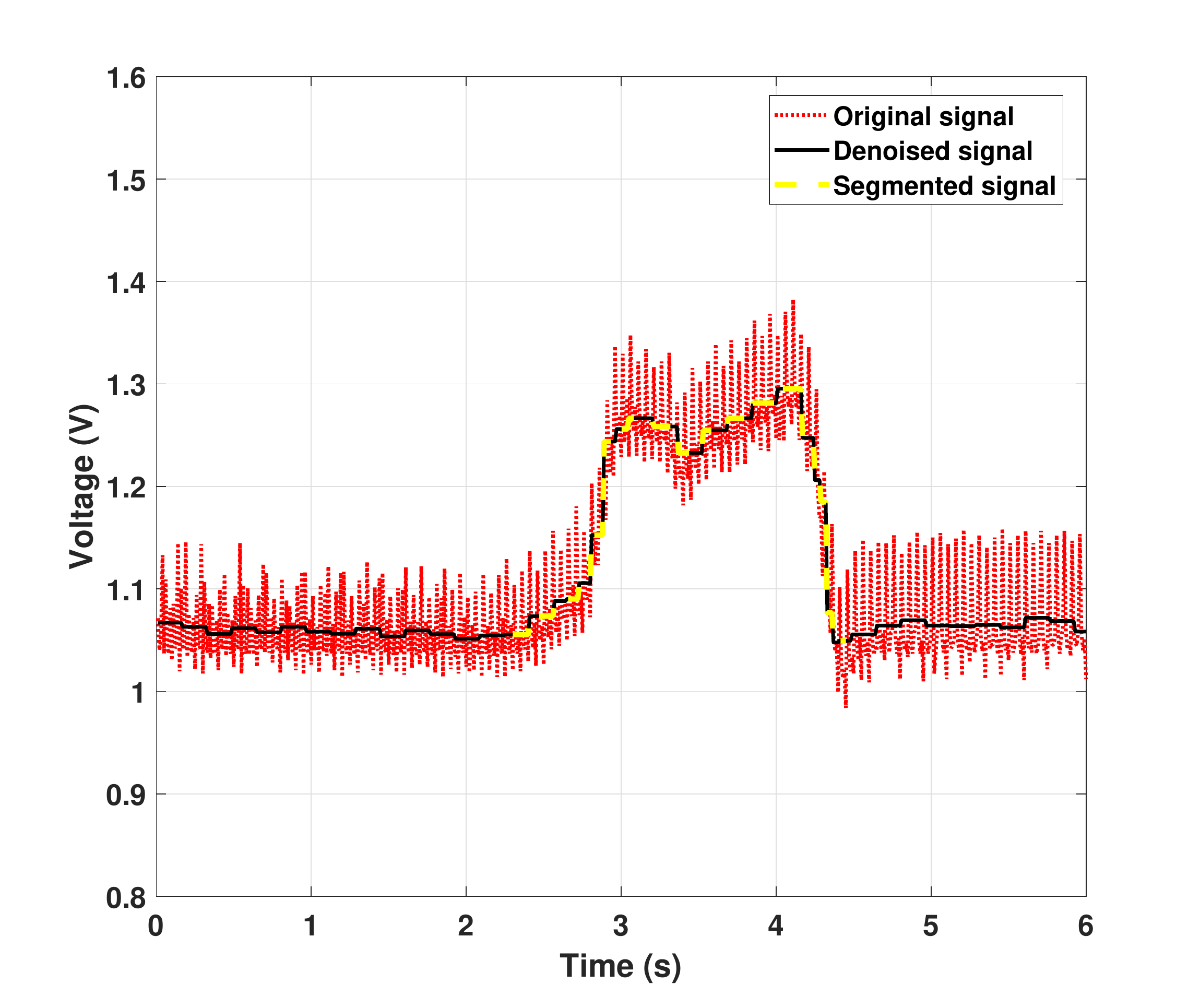}
\centering
\caption{The effect of Discrete Wavelet Denoising and Simple Thresholding blocks on the time domain from received visible light signal at 20~cm.}
\label{fig:DWD_Simple_Thresholding_effect}
\vspace{-4mm}
\end{figure}

Determining the proper threshold value is an important detail in the denoising process.  A large threshold may over-smooth the recovered signal, losing important time-domain detail information.  A threshold that is too small cannot effectively eliminate the noise.  An appropriate threshold should be selected to balance these competing requirements. Donoho and Johnstone  \cite{thhhh} have done a large amount of research in this area. There are two main categories: global and level dependent thresholding. Global thresholding employs the same threshold value for all coefficients at every decomposition level. In level dependent thresholding, it is necessary to find a suitable and possibly different threshold for each decomposition level.
In \cite{th3}, four thresholding techniques were analyzed and evaluated for denoising performance, based on the calculation of mean square error: \texttt{rigrsure}, \texttt{heursure}, \texttt{sqtwolog} and \texttt{minimaxi}.  We found that the \texttt{rigrsure} showed an optimum performance in our case. As shown in Fig.~\ref{fig:noise_p}, the red wave with \texttt{rigrsure} principle is the smoothest one and provides the foundation of subsequent operations. 
The denoising effectiveness is demonstrated with another signal in Fig.~\ref{fig:DWD_Simple_Thresholding_effect}.

\vspace{2mm}
\noindent\textbf{Gesture Detection}

Since gestures generally lasted only 2-3 seconds, much of the 6 seconds worth of collected data could be discarded.  We therefore needed to detect the beginning and end for every gesture, in between which the received signal had large changes in magnitude and shape due to hand movement.  Only data between these points were forwarded for classification.  Generally, the reflected intensity becomes much larger when the user is performing a gesture.  Therefore, we used a thresholding scheme to detect the start and end points of each gesture \emph{after} the denoising operation. Note that this is now a time-domain thresholding, entirely separate from the thresholding operations used in denoising.   We found that suitable thresholds are slightly greater ($\sim 10$\%) than the averaged measured signal intensity \emph{before the gesture begins}. 
The suitable thresholds were obtained based on the received signal value without any initial hand movements because the fluctuation of waveform could be very tiny as shown in Fig.~\ref{fig:Gesture_set}(d) while sweeping from right to left. If we chose the thresholds greater than 10\%, the details like this would be removed. With this threshold, we did not lose any gesture related data, but got rid of redundant data. Thus, it was the optimal option for the recognition accuracy.
The absolute scale of this threshold is variable based on the environmental situation and signal intensities observed during the measurement. For example, the average background intensity measured in a light room is larger than in a dark room.  An example of this gesture detection process is shown on measured data  by the amber, dashed curve in Fig.~\ref{fig:DWD_Simple_Thresholding_effect}.

\begin{figure}[!t]
\centering
\includegraphics[width=3.4in]{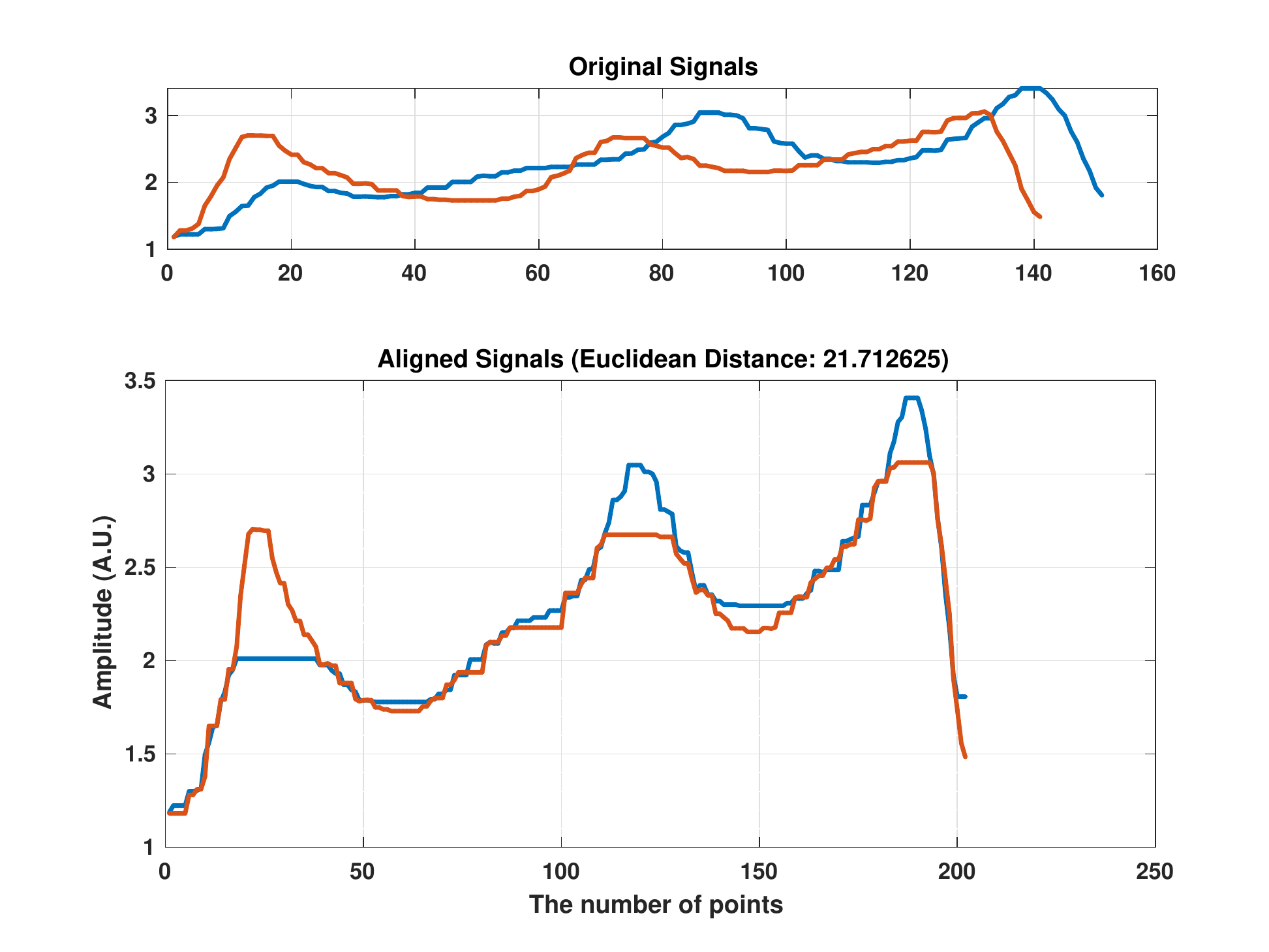}
\caption{The DTW result for two samples of infrared light signal at 20~cm with different duration from the same gesture.}
\label{dtws}
\end{figure}

\begin{figure}[!t]
\centering
\includegraphics[width=3.4in]{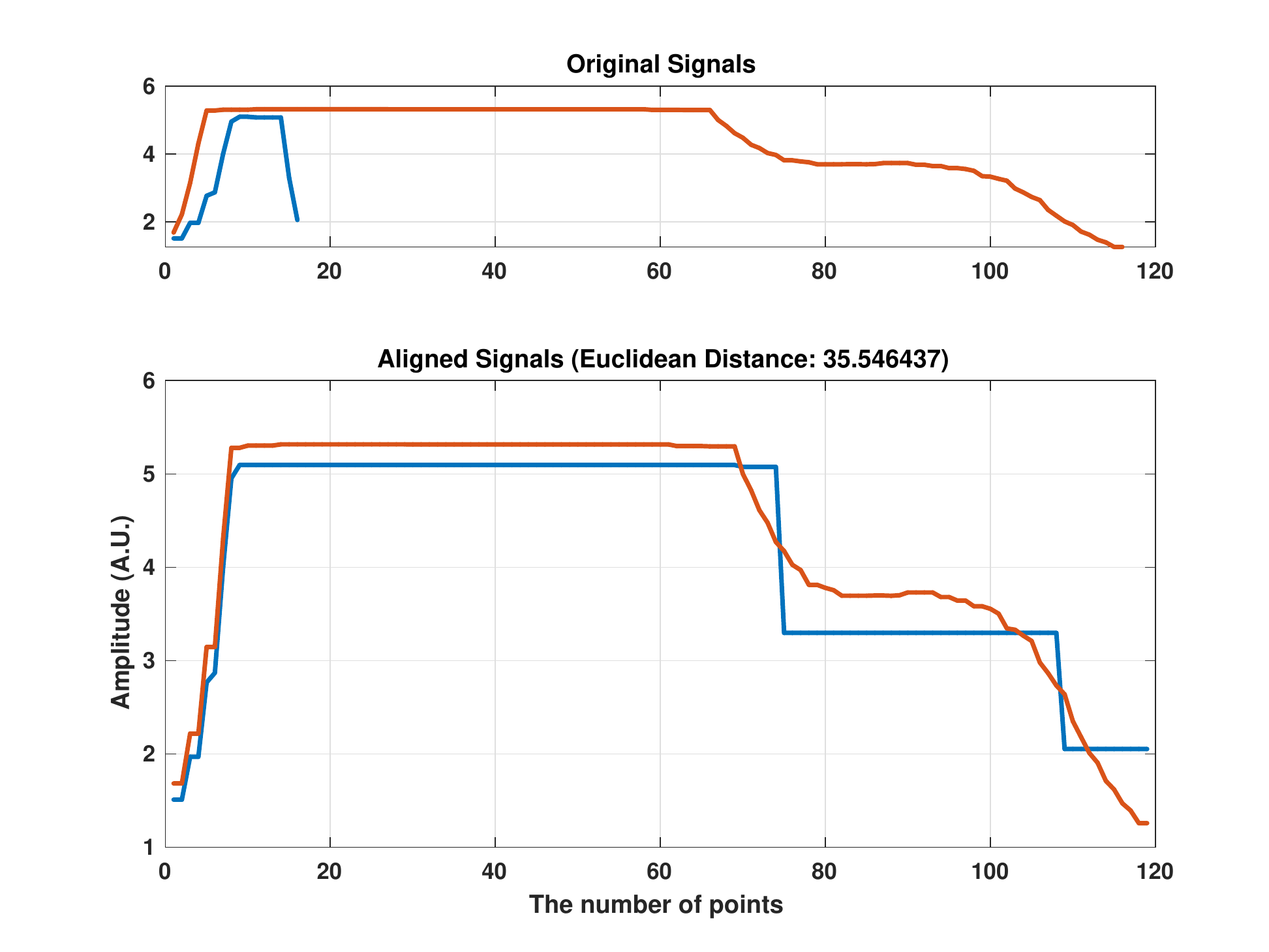}
\caption{ The DTW result for two samples of infrared light signal at 20~cm with different duration from different gesture with similar wave shape.}
\label{dtw}
\end{figure}

After deleting the unusable data, we obtain a time series of the gesture with an unpredictable number of data points. Obviously, the filtered signals will not always have the same length at this stage.  However, the follow-on classification algorithms require that all the input data have a vectorized representation of the \emph{same} length, which allows the algorithms to efficiently execute matrix operations in batch. Signals can be time-scaled or zero-padded to cause their vector lengths to match.   Dynamic Time Warping (DTW) is a method that aligns two time series to measure the similarities between them \cite{dtw_}. For example, two signals for a same gesture with different performance speed have similar shapes but different magnitudes or lengths. The DTW can make them match perfectly. We can get two new time series of same length after applying the DTW. Meanwhile, we find that it makes two waves with different duration from a same gesture match very well as shown in Fig.~\ref{dtws}. This approach adjusts the shape of waves to an identical one for one gesture to eliminate the diversity from various users. This can be an advantage in the recognition process. However, it reduces the distinctiveness for different gestures which have similar wave shapes because it aligns all the waves in a similar shape as shown in Fig.~\ref{dtw}. In this figure, those two waves are from two different gestures with similar wave shapes. As can be seen, the shape of the blue wave is changed with the red one. The dissimilarity among different gestures is reduced through this procedure. This operation will make all the waves, no matter whether they come from same gestures or different gestures, have a similar shape, resulting in error classification. Furthermore, it will damage the recognition accuracy significantly.
Therefore, we found that the most reliable gesture recognition was accomplished by employing zero-padding to match their lengths.

\begin{figure}[t]
\includegraphics[width=0.5\textwidth,,height=6cm]{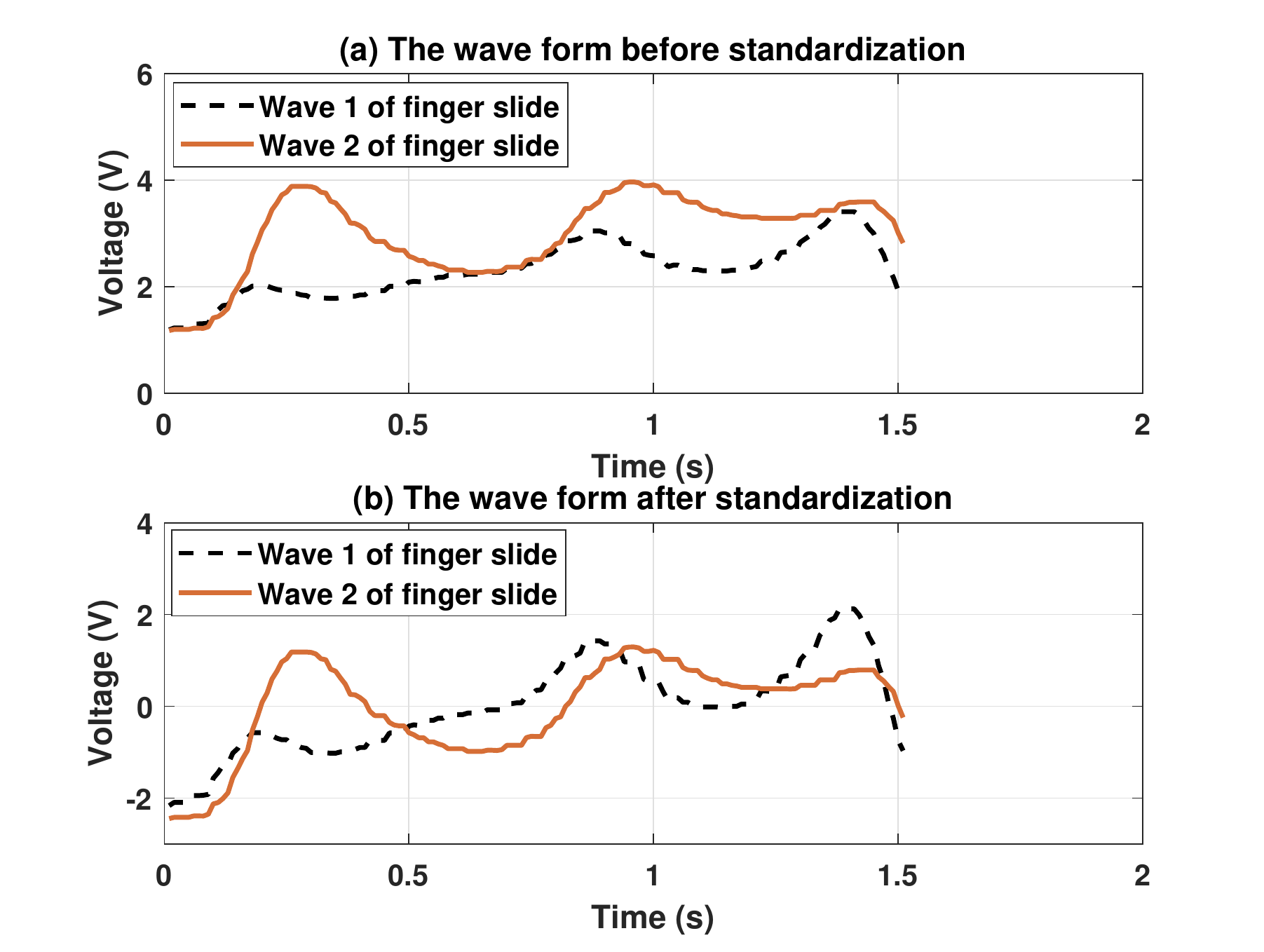}
\centering
\caption{The effect of Z scores normalization block on the time domain signal from received infrared light signal at 20~cm. }
\label{fig:standardization_effect}
\vspace{-4mm}
\end{figure}

\vspace{2mm}
\noindent\textbf{Standardization}

Due to the diversity of factors involved when different users perform gestures, the magnitudes of measured light signals from the \emph{same} gesture may be highly variable, even though their wave shapes are generally similar. For example, the distance $d$ between a user's hand and the receiver affects the magnitude of the reflected signal.  These variations deteriorate the recognition accuracy. An example is shown in Fig.~\ref{fig:standardization_effect}(a).  Two denoised signal plots are shown, one each of the same gesture from two different volunteers.  While the waves have similar shapes, they have quite different magnitudes.  To solve this problem, the denoised signals must be made to have similar magnitudes.  This can be done with the Z scores method, which compares the Standard Deviations (SD) of the compared signals.  By visual inspection of the signals in Fig.~\ref{fig:standardization_effect}(a), the one with a larger magnitude also has a larger standard deviation.  Signals can therefore be scaled by their standard deviation to get similar magnitudes. Z scores is a standardization method that simply converts a data set to a distribution with zero mean and unity standard deviation.  As shown in the Fig.~\ref{fig:standardization_effect}(b), the two previously described waves are shown again, after applying the Z scores standardization; they now exhibit similar magnitudes. This method made our system agnostic to different users and measurement conditions.

\subsection{Classifier Training}      
\begin{figure*}[htbp]
\centering
\includegraphics[width=1.8\columnwidth]{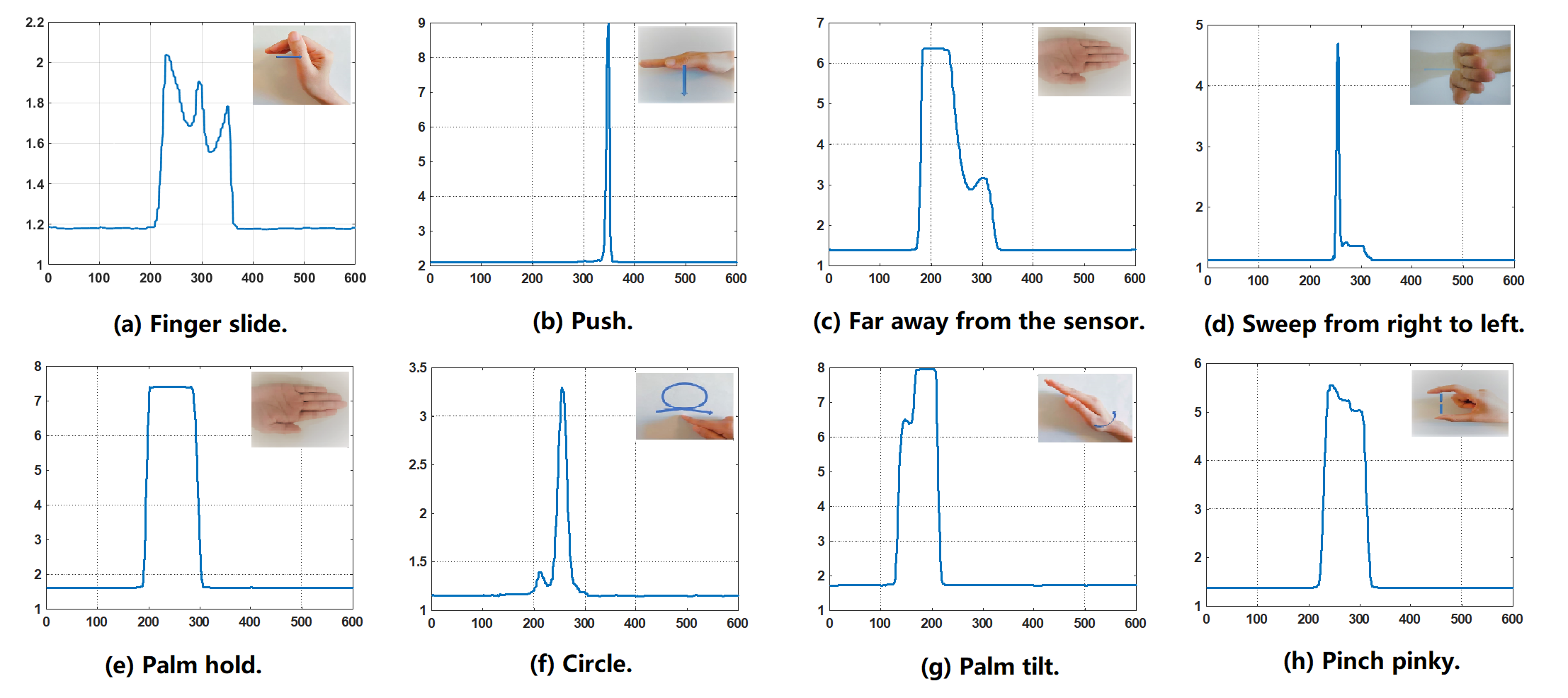}
\caption{{The time-series waves of received infrared light signal at 20~cm}  with the corresponding gesture.}
\label{fig:Gesture_set}
\end{figure*}

After the signal processing is completed, all of the data sets were used as feature vectors in a training set to build a gesture classification method using the K-nearest-neighbors method (KNN). %\footnote
We applied KNN and SVM in our study. The recognition accuracy showed that the KNN behaved better than SVM in the recognition of 8 gestures. Based on the data analysis, the variances of our feature vectors were small. Due to the tiny difference and much overlap of all the waves (as shown in Fig.~\ref{fig:Gesture_set}), KNN was found to be more suitable for our proposed system.
Hand gestures were selected {based on the common HCI tasks}\cite{hand1,light16}.  The waves shown in Fig.~\ref{fig:Gesture_set} are the {denoised}  time-domain signals. 
After data analysis, we found that the variance of our feature vectors was small. This could also be observed visually as relatively small differences in the 8 gestures in the time-domain waves.  KNN is a non-parametric lazy method that utilizes the distance between each sample to separate them into several classes without any assumptions in the classification and regression applications \cite{knn11}.  Therefore, due to the small difference and strong overlap of all the waves, KNN was found to be more suitable for our method.% 

To build a training model, we collected several waves for each gesture.  As mentioned earlier, statistical results showed that the volunteers  completed the gesture movements shown in Fig.~\ref{fig:Gesture_set} in about 2-3 seconds.  When the data collection was completed, the raw received data was denoised, trimmed, normalized, and padded as required and as discussed previously.  Using these data, all the feature vectors were used to train a KNN classification model and get the best parameters for our gesture recognition system. {New waves with accurate record of gesture type were used to test the trained model.}  

\section{Evaluation, Results and Discussion}
\label{sec:Results}
In this section, we evaluate our gesture recognition system using the collected data from real volunteers with K-fold validation. First, we discuss the data collection method and software used. Then we describe a K-fold validation procedure used to evaluate the gesture classification performance with less bias. We then present a calculated confusion matrix that classifies the accuracy for eight different gestures and visualizes the classification error. Further, we present analysis of the gesture classification accuracy when using different light sources at different distances. Finally, the impact of environment lighting conditions in gesture classification accuracy is presented.

\subsection{Data Collection}
Referring to the setup shown in Fig.~\ref{fig:Hardware_setup}  volunteers made hand gestures in front of the light source.  For each gesture, the 600 points were saved in the Raspberry Pi. {The data collection used  Python scripting language run in PiPlate and Raspberry Pi.  The signal processing and classification algorithm using Matlab was applied in the PC offline based on the saved data in Raspberry Pi.}

 { We instructed volunteers how to perform each gesture, and gave them several minutes to practice until they were comfortable with each gesture.  We recorded data from 5 volunteers (two females of 25 and 27 years old and three males, two of 27 years old and one of 24 years old.) performing 24 repetitions of the 8 gestures (960 waves) shown in} Fig.~\ref{fig:Gesture_set} {at distance 20 cm from the sensor for infrared source in normal indoor room lighting conditions. There was only one window in the room and the curtains were always closed. The color of the wall was white. The light of environment was from fluorescent lights in the ceiling and computer screen. The background color, material characteristics, distance from the gesture scene etc. are not expected to impact the results significantly. The background will affect only the initial received signal value when there is no hand movement. The signal that was useful for gesture identification was the reflected light by hand movement between the photodetector and the background. When volunteers performed the gesture, they could sit in front of the photodetector or stand by the side of the photodetector, and then perform the gesture in front of the photodetector at the specific distance. The received signals of those two situations would have small difference in the reflected intensity.
 To observe the impact of ambient light, the same number of waves  were recorded at distance 20 cm infrared  sensing in the dark indoor room (no lighting) conditions.  Further, to obtain the effect of sensing distance on accuracy, the same number of waves were collected at distance 35 cm from the sensor for infrared sources in normal indoor room lighting conditions. There were 3 data sets with same number of waves for infrared sensing. The same 3 data sets under same conditions were collected using the visible light source to determine the effect of different sources on gesture recognition performance.}

\subsection{Gesture Recognition Accuracy}

\begin{figure}[t]
\includegraphics[width=0.48\textwidth,height=6cm]{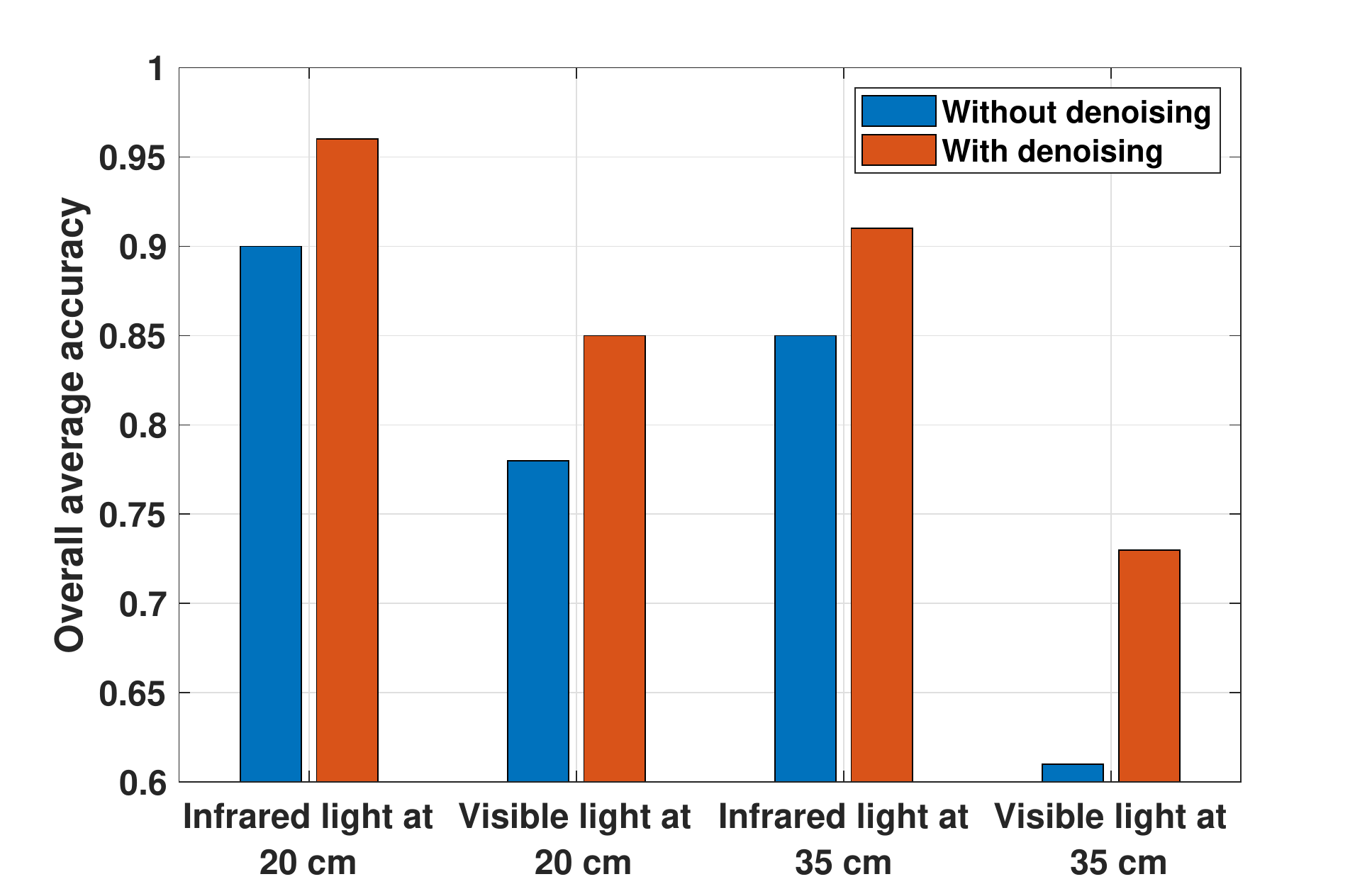}
\centering
\caption{Effect of denoising step on the average accuracy for both infrared and visible light source at 20 and 35~cm (ambient light is on).}
\label{fig:Accurcy_denoising_Effect}
\vspace{-4mm}
\end{figure}

\begin{table*} [!t]
\centering
    \begin{tabular}{c c c c c c c c}  
 \hline
 
\textbf{Distance (cm) }  &\textbf{5}&\textbf{10}&\textbf{15}&\textbf{20}&\textbf{25}&\textbf{30}&\textbf{35} \\ 
 \hline
\textbf{Measured radiated power
(mW)}& 15.6&5.9&3.05&1.86&1.22&0.87&0.65\\ 

\textbf{Measured power at detector reflected from a flat hand 
(mW)}& 1.09& 0.415&0.248&0.193&0.065&0.053&0.048\\ 
 
\hline
 \end{tabular}
 \caption{Power measurements for infrared light setup.}
 \label{table:infpower}
\end{table*}

\begin{table*} [!t]
\centering
    \begin{tabular}{c c  c c c c c c}  
 \hline
  \textbf{Distance (cm) }  &\textbf{5}&\textbf{10}&\textbf{15}&\textbf{20}&\textbf{25}&\textbf{30}&\textbf{35} \\ 
  \hline
\textbf{Power measured by power meter(mW)}& 3.02&1.46&0.750&0.460&0.320&0.240&0.190\\ 
 
\textbf{Reflected power measured by power meter (mW)}& 0.0512& 0.0412&0.0364&0.0337&0.0315&0.0296&0.0286\\ 
 
\hline
 \end{tabular}
 \caption{Power measurements for visible light setup.}
 \label{table:visiblepower}
\end{table*}
In this section, we evaluate the accuracy of our system at different distances, with different sources, and in different ambient lighting conditions. {All the results were obtained with 10-fold cross validation to use all the data and make the parameters fine-tuned.  In 10-fold cross validation, all the waves were divided into 10 equal size subsets randomly. One of the 10 subsets was taken as testing data, the remaining 9 subsets were used as training data. Then the cross validation was repeated 10 times, and each of the 10 subsets was used as the testing data only once.
Note that the data sets were divided randomly, and the model was not supposed to be subject-dependent. The training data sets were overlapped with each other. From our results, the testing accuracy results of 10 models were not significantly different (the SD range of those 10 matrices were within 5\%).
Finally, 10 confusion matrices were averaged to obtain a final result.
To measure the reproducibility of the system, the accuracy result of leave-one-subject-out validation was calculated using the same samples. The confusion matrix of 10-fold cross and leave-one-subject-out validation results for infrared light sensing around 20~cm are shown in Table~\ref{inf1}  and Table~\ref{inf1_leave}. The overall accuracy of cross validation result was 96.13\% (SD = 2.59\%). And the overall recognition result of leave-one-subject-out validation was 92.13\% (SD = 3.14\%). This result represented a more realistic status compared with cross-validation result.}

\begin{table}[t]
\centering
\begin{tabular}{@{}clcccccccc@{}}
\toprule
\multicolumn{1}{l}{}                                                                                 & \multicolumn{9}{c}{Estimating Gesture}                                                                                                                                                                                                           \\ \midrule
\multicolumn{1}{|c|}{\multirow{9}{*}{\begin{tabular}[c]{@{}c@{}}\rotatebox{90}{Performing Gesture\hspace{1cm}}\end{tabular}}} & \multicolumn{1}{l|}{}    & \multicolumn{1}{l|}{(a)} & \multicolumn{1}{l|}{(b)} & \multicolumn{1}{l|}{(c)} & \multicolumn{1}{l|}{(d)} & \multicolumn{1}{l|}{(e)} & \multicolumn{1}{l|}{(f)} & \multicolumn{1}{l|}{(g)} & \multicolumn{1}{l|}{(h)} \\ \cmidrule(l){2-10} 
\multicolumn{1}{|c|}{}                                                                               & \multicolumn{1}{l|}{(a)} & \textbf{0.93}            & 0                        & 0.04                     & 0                        & 0                        & 0.01                     & 0                        & 0.01                     \\ \cmidrule(lr){2-2}
\multicolumn{1}{|c|}{}                                                                               & \multicolumn{1}{l|}{(b)} & 0             & \textbf{0.92}            & 0                        & 0.06                     & 0                        & 0.02                     & 0                        & 0                        \\ \cmidrule(lr){2-2}
\multicolumn{1}{|c|}{}                                                                               & \multicolumn{1}{l|}{(c)} & 0                        & 0                        & \textbf{1}               & 0                        & 0                        & 0                        & 0                        & 0                        \\ \cmidrule(lr){2-2}
\multicolumn{1}{|c|}{}                                                                               & \multicolumn{1}{l|}{(d)} & 0.01                     & 0.01                     & 0                        & \textbf{0.97}            & 0                        & 0.01                     & 0                        & 0                        \\ \cmidrule(lr){2-2}
\multicolumn{1}{|c|}{}                                                                               & \multicolumn{1}{l|}{(e)} & 0                        & 0                        & 0                        & 0                        & \textbf{0.98}            & 0                        & 0.02                     & 0                        \\ \cmidrule(lr){2-2}
\multicolumn{1}{|c|}{}                                                                               & \multicolumn{1}{l|}{(f)} & 0.01                     & 0                        & 0                        & 0.03                     & 0                        & \textbf{0.96}            & 0                        & 0                        \\ \cmidrule(lr){2-2}
\multicolumn{1}{|c|}{}                                                                               & \multicolumn{1}{l|}{(g)} & 0                        & 0                        & 0.03                     & 0                        & 0                        & 0                        & \textbf{0.97}            & 0                        \\ \cmidrule(lr){2-2}
\multicolumn{1}{|c|}{}                                                                               & \multicolumn{1}{l|}{(g)}                      & 0.02                     & 0                        & 0                        & 0.01                     & 0                        & 0.01                     & 0                        & \textbf{0.96}            \\ \bottomrule
\end{tabular}
\caption{The 10-fold validation confusion matrix of infrared light-wave sensing around 20~cm (ambient light is on).}
\label{inf1}
\end{table}

\begin{table}[t]
\centering
\begin{tabular}{@{}clcccccccc@{}}
\toprule
\multicolumn{1}{l}{}                                                                                 & \multicolumn{9}{c}{Estimating Gesture}                                                                                                                                                                                                           \\ \midrule
\multicolumn{1}{|c|}{\multirow{9}{*}{\begin{tabular}[c]{@{}c@{}}\rotatebox{90}{Performing Gesture\hspace{1cm}}\end{tabular}}} & \multicolumn{1}{l|}{}    & \multicolumn{1}{l|}{(a)} & \multicolumn{1}{l|}{(b)} & \multicolumn{1}{l|}{(c)} & \multicolumn{1}{l|}{(d)} & \multicolumn{1}{l|}{(e)} & \multicolumn{1}{l|}{(f)} & \multicolumn{1}{l|}{(g)} & \multicolumn{1}{l|}{(h)} \\ \cmidrule(l){2-10} 
\multicolumn{1}{|c|}{}                                                                               & \multicolumn{1}{l|}{(a)} & \textbf{0.87}            & 0                        & 0.01                     & 0.01                       & 0                        & 0.02                     & 0.02                        & 0.07                     \\ \cmidrule(lr){2-2}
\multicolumn{1}{|c|}{}                                                                               & \multicolumn{1}{l|}{(b)} & 0             & \textbf{0.89}            & 0                        & 0.04                     & 0                        & 0.02                   & 0                        & 0.05                        \\ \cmidrule(lr){2-2}
\multicolumn{1}{|c|}{}                                                                               & \multicolumn{1}{l|}{(c)} & 0                        & 0                        & \textbf{0.95}               & 0                        & 0                        & 0                        & 0                        & 0.5                        \\ \cmidrule(lr){2-2}
\multicolumn{1}{|c|}{}                                                                               & \multicolumn{1}{l|}{(d)} & 0.01                     & 0.03                     & 0                        & \textbf{0.93}            & 0                        & 0.03                     & 0                        & 0                        \\ \cmidrule(lr){2-2}
\multicolumn{1}{|c|}{}                                                                               & \multicolumn{1}{l|}{(e)} & 0                        & 0.01                        & 0                        & 0.05                        & \textbf{0.91}            & 0                        & 0.02                     & 0.01                        \\ \cmidrule(lr){2-2}
\multicolumn{1}{|c|}{}                                                                               & \multicolumn{1}{l|}{(f)} & 0.02                     & 0.01                        & 0                        & 0.01                     & 0                        & \textbf{0.94}            & 0.01                       & 0.01                        \\ \cmidrule(lr){2-2}
\multicolumn{1}{|c|}{}                                                                               & \multicolumn{1}{l|}{(g)} & 0                        & 0.01                        & 0.01                     & 0                        & 0                      & 0                        & \textbf{0.96}            & 0.02                        \\ \cmidrule(lr){2-2}
\multicolumn{1}{|c|}{}                                                                               & \multicolumn{1}{l|}{(g)}                      & 0.02                     & 0                        & 0.02                        & 0.03                     & 0                        & 0.01                     & 0                        & \textbf{0.92}            \\ \bottomrule
\end{tabular}
\caption{The leave-one-subject-out validation confusion matrix of infrared light-wave sensing around 20~cm (ambient light is on).}
\label{inf1_leave}
\end{table}

\begin{table}[t]
\centering
\begin{tabular}{@{}clcccccccc@{}}
\toprule
\multicolumn{1}{l}{}                                                                                 & \multicolumn{9}{c}{Estimating Gesture}                                                                                                                                                                                                           \\ \midrule
\multicolumn{1}{|c|}{\multirow{9}{*}{\begin{tabular}[c]{@{}c@{}}\rotatebox{90}{Performing Gesture\hspace{1cm}}\end{tabular}}} & \multicolumn{1}{l|}{}    & \multicolumn{1}{l|}{(a)} & \multicolumn{1}{l|}{(b)} & \multicolumn{1}{l|}{(c)} & \multicolumn{1}{l|}{(d)} & \multicolumn{1}{l|}{(e)} & \multicolumn{1}{l|}{(f)} & \multicolumn{1}{l|}{(g)} & \multicolumn{1}{l|}{(h)} \\ \cmidrule(l){2-10} 
\multicolumn{1}{|c|}{}                                                                               & \multicolumn{1}{l|}{(a)} & \textbf{0.81}  &       0      &   0   & 0.02  &  0 &   0.09 &   0.09  &  0                   \\ \cmidrule(lr){2-2}
\multicolumn{1}{|c|}{}                                                                               & \multicolumn{1}{l|}{(b)} & 0  &   \textbf{0.91}  &  0.03     &    0.04  &  0.02    &     0  &  0   & 0                      \\ \cmidrule(lr){2-2}
\multicolumn{1}{|c|}{}                                                                               & \multicolumn{1}{l|}{(c)} & 0 &   0.00 &   \textbf{0.85}   &    0   & 0.02&     0     &    0.05  &  0.08                        \\ \cmidrule(lr){2-2}
\multicolumn{1}{|c|}{}                                                                               & \multicolumn{1}{l|}{(d)} &    0.01  &       0.02   & 0.02  &  \textbf{0.86}  &  0.01  & 0.01   & 0.05 &  0.04                      \\ \cmidrule(lr){2-2}
\multicolumn{1}{|c|}{}                                                                               & \multicolumn{1}{l|}{(e)}  &0   & 0 &  0.08 & 0.01  &  \textbf{0.85}  &      0  &  0.02  &  0.05                    \\ \cmidrule(lr){2-2}
\multicolumn{1}{|c|}{}                                                                               & \multicolumn{1}{l|}{(f)} &    0.03   &     0  &       0     &    0  &       0  &   \textbf{0.83}  &  0.12  &  0.02                  \\ \cmidrule(lr){2-2}
\multicolumn{1}{|c|}{}                                                                               & \multicolumn{1}{l|}{(g)} &  0.02    &    0   &      0.03   & 0  &  0.01 &  0.09  &   \textbf{0.83} &  0.02                       \\ \cmidrule(lr){2-2}
\multicolumn{1}{|c|}{}                                                                               & \multicolumn{1}{l|}{(h)}                      &  0.01    &    0   & 0.06  &  0 &  0.05  &  0.01  &  0.03 &  \textbf{0.84}           \\ \bottomrule
\end{tabular}
 \caption{The confusion matrix of visible light-wave sensing around 20 cm (ambient light is on).}  
 \label{visible1}
 
 \end{table}

\subsubsection{Accuracy of infrared and visible light-wave sensing compared at the same distance ($d =20$~cm and $d=35$~cm)}

The data set of infrared light-wave sensing had  960 gesture waves named as infrared data set around 20~cm. The infrared data set around 35~cm had the same number of waves.
The forms of the data sets for visible light-wave sensing at 20~cm and 35~cm were identical.

{As shown in }Table~\ref{inf1} and Table~\ref{visible1}, the accuracy of gesture recognition at $d=20$~cm is with using the infrared light than that with the visible light.

In order to quantify this difference, we measured the light power as a function of distance from the source using a calibrated optical power meter.  Using  Table~\ref{table:infpower} and Table~\ref{table:visiblepower}, ``measured radiated power'' refers to power measured directly in front of the light source at different distances.  ``Measured power at detector reflected from flat hand'' refers to the power measured after reflection from a hand held flat and normal to the source beam.  In this case, the power meter probe was placed directly in front of the usual system photodetector.
The distance refers to where the hand was held during the power measurement.
{One can observe from the tables that the infrared light power is much larger than the visible light power. As shown in }Fig.~\ref{fig:inf_}, {the wavelength of our visible light source is from 400~nm to 900~nm, and the wavelength of our infrared light source is 940~nm. The spectrum was obtained using an Ocean Insight USB2000+ spectrometer~\cite{spectro}. During measurements, the light sources, both visible and infrared, were pointed directly at the tip of the optical fiber that conveys light to the spectrometer. The distance between the light sources and fiber tip, as well as pointing of the fiber tip, were adjusted to avoid saturating the spectrometer's detector. Spectral data were stored by the spectrometer control software during acquisition.
\begin{figure}[t]
\includegraphics[width=1\columnwidth,height=6cm]{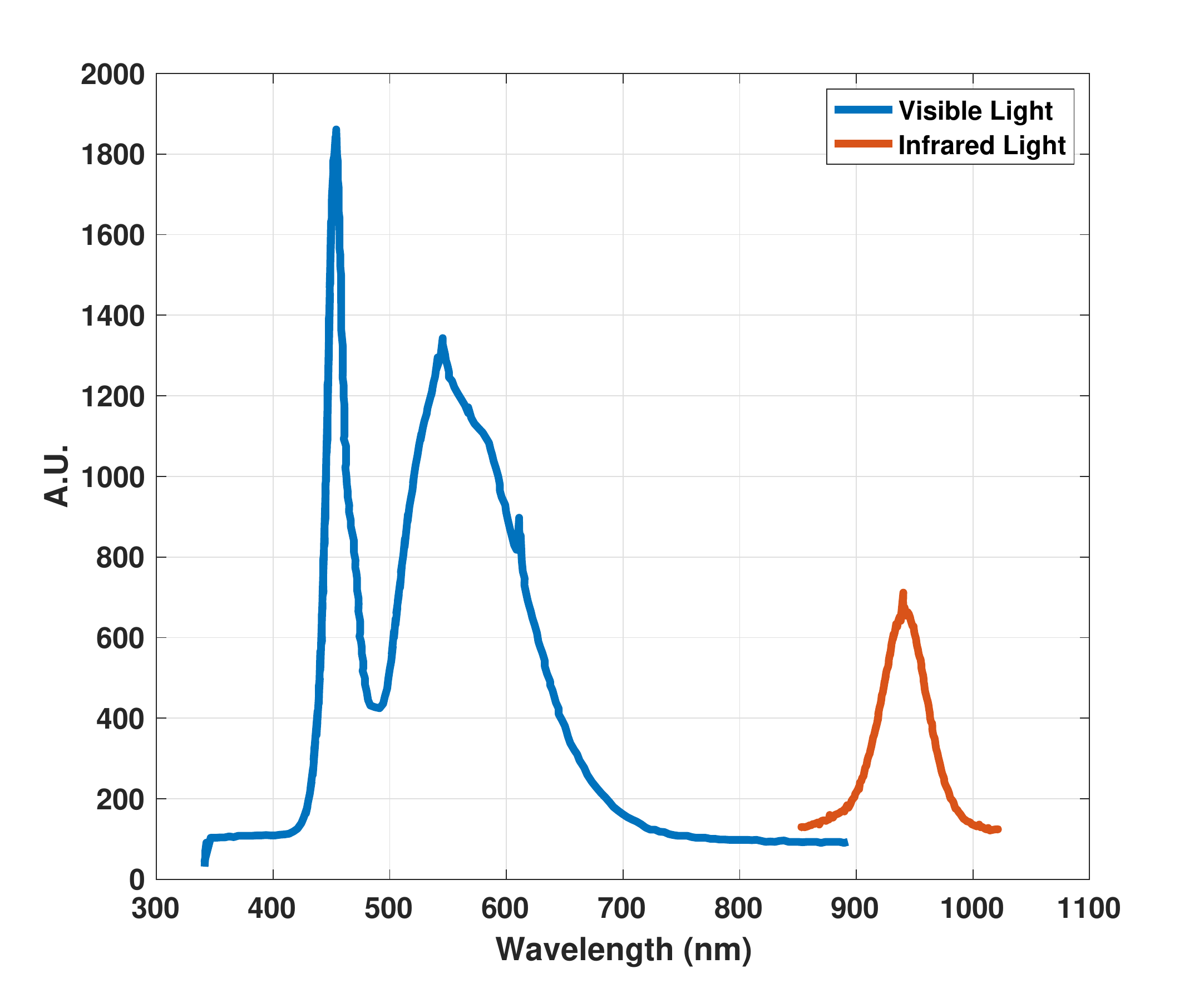}
\centering
\caption{The measured spectrum of our infrared and visible light sources (The different peak levels of infrared and visible light are not necessarily indicative of relative light intensity. The visible and infrared spectra were scaled separately, and then combined in the figure plot). }
\label{fig:inf_}
\end{figure}

\begin{figure}[t]
\includegraphics[width=1\columnwidth,height=6cm]{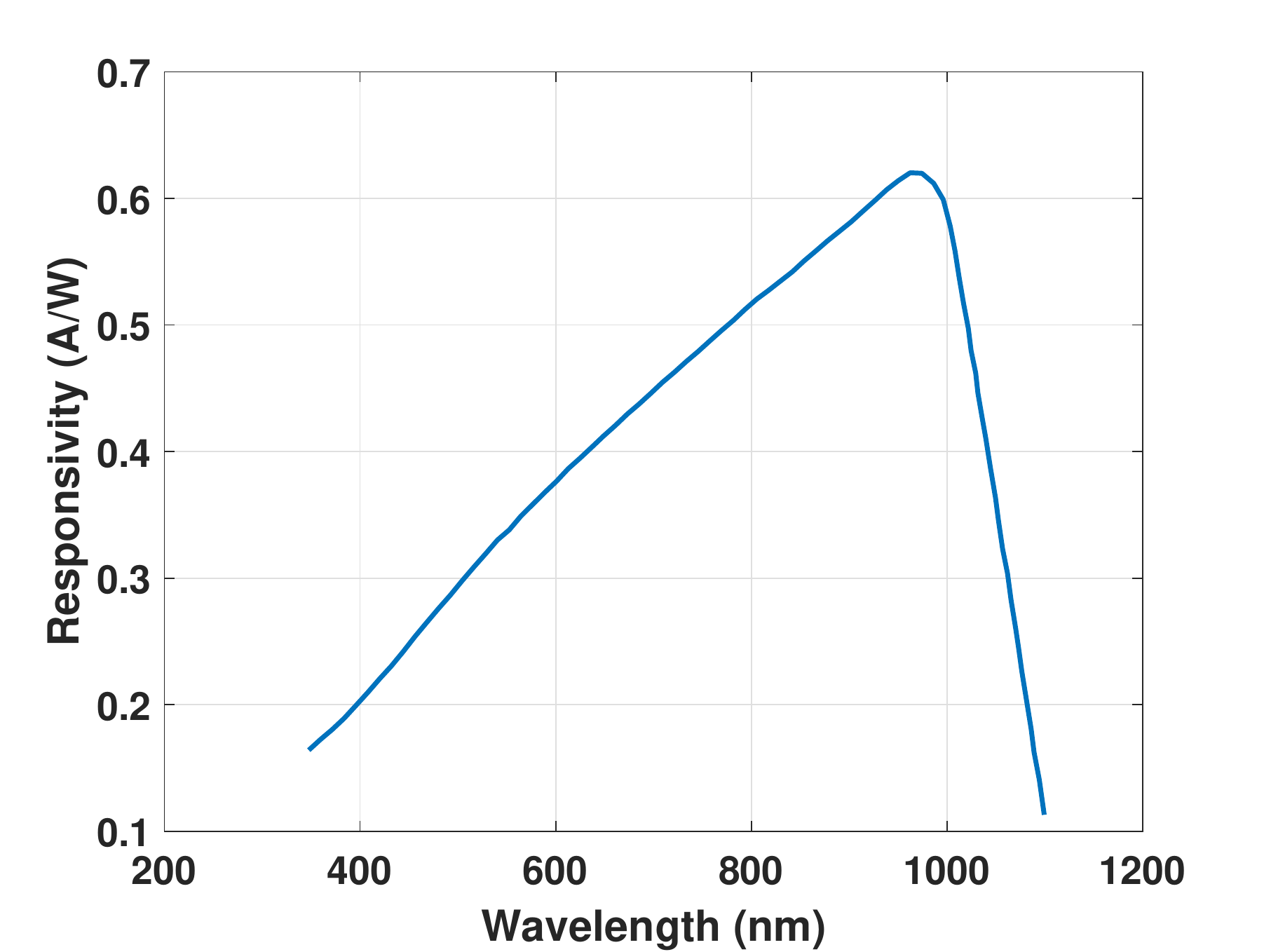}
\centering
    \caption{PDA100A spectrum responsivity \cite{pda}. }
\label{fig:response_}
\vspace*{-4mm}
\end{figure}
The spectrum reponsivity of the photodector is shown in }Fig.~\ref{fig:response_} {which is regenerated from the PDA100A manual} \cite{pda}. {Because of the larger light power and reponsivity of the photodector, the infrared light  generates larger voltages at the detector.}
The larger voltage signals generally seem to raise the signal-to-noise ratio (SNR) of the system, resulting in higher recognition precision. The results in  Fig.~\ref{fig:visible_comparison} and Fig.~\ref{fig:infrared_comparison} at $d=35$~cm reveal the same conclusion.  %(\hl{Figures 12 and 13 should have the same vertical scale.})

\subsubsection{Accuracy at different distances}       
\begin{figure}[t]
\includegraphics[width=0.48\textwidth,height=6cm]{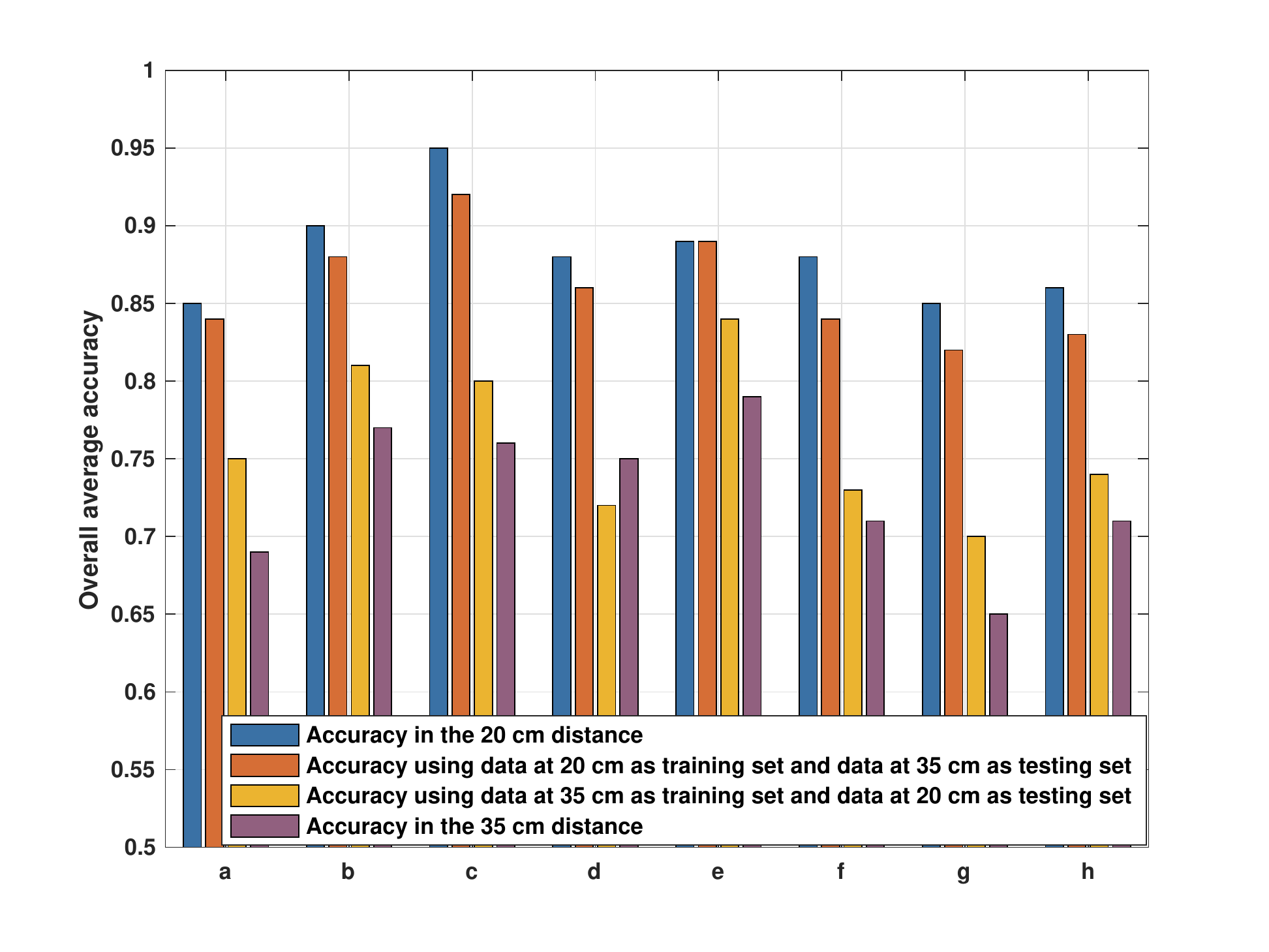}
\centering
\caption{Accuracy results of visible light-wave sensing at 20 and 35~cm (ambient light is on; mean = 84.75\%, SD = 2.96\% for the accuracy at 20~cm; mean = 86.00\%, SD = 3.42\% for the accuracy using the data set at 20~cm as training set and data set at 35~cm as testing set; mean = 76.13\%, SD = 4.94\% for the accuracy using the data set at 35~cm as training set and data set at 20~cm as testing set; mean = 71.13\%, SD = 4.91\% for the accuracy at 20~cm). }
\label{fig:visible_comparison}
\vspace{-2mm}
\end{figure}

\begin{figure}[t]
\includegraphics[width=0.48\textwidth,height=6cm]{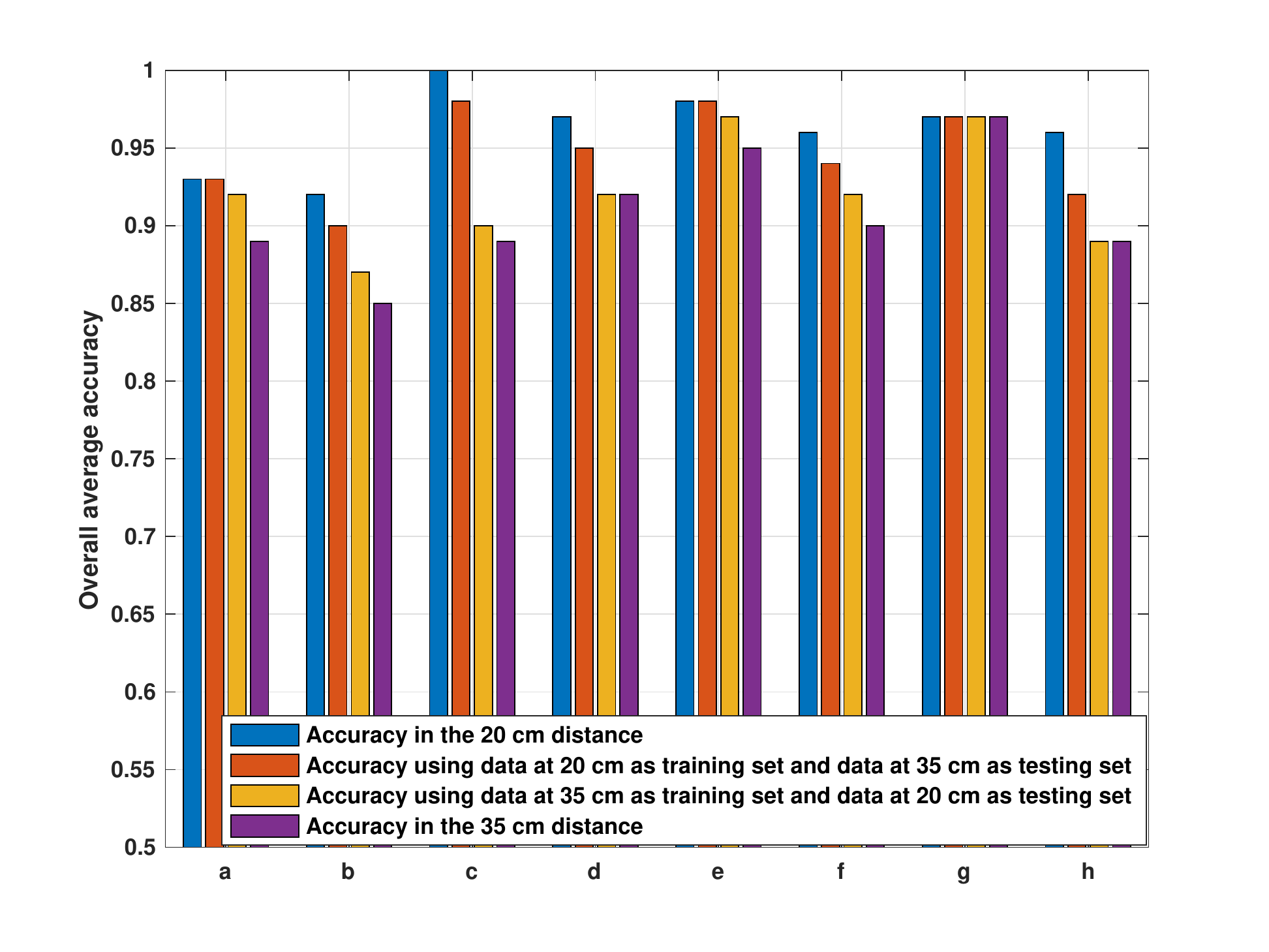}
\centering
\caption{Accuracy results of infrared light-wave sensing at 20 and 35~cm (ambient light is on; mean = 96.13\%, SD = 2.59\% for the accuracy at 20 cm; mean = 94.62\%, SD = 2.92\% for the accuracy using the data set at 20~cm as training set and data set at 35~cm as testing set; mean = 92.00\%, SD = 3.55\% for the accuracy using the data set at 35~cm as training set and data set at 20~cm as testing set; mean = 84.75\%, SD = 2.96\% for the accuracy at 20~cm).}
\label{fig:infrared_comparison}
\vspace{-2mm}
\end{figure}

Fig.~\ref{fig:visible_comparison} and Fig.~\ref{fig:infrared_comparison} reveal the effect of distance on gesture recognition accuracy.  Accuracy decreases with increasing distance due to the lower reflected light intensity.   The reduced accuracy at increased distance with both infrared and visible light sources can also be attributed to lower light intensities reducing the overall system SNR. 
The results of 20 cm sensing data for training with 30 cm data for testing and vice versa are shown in Fig.~\ref{fig:visible_comparison} and  Fig.~\ref{fig:infrared_comparison} , to show the sensitivity on different training and testing distances. The effect of distance for collecting testing data had negligible impact on the accuracy of the system. The accuracies of the same training data set with different testing sets were found to be similar.

\begin{figure}[t]
\includegraphics[width=0.48\textwidth,height=6cm]{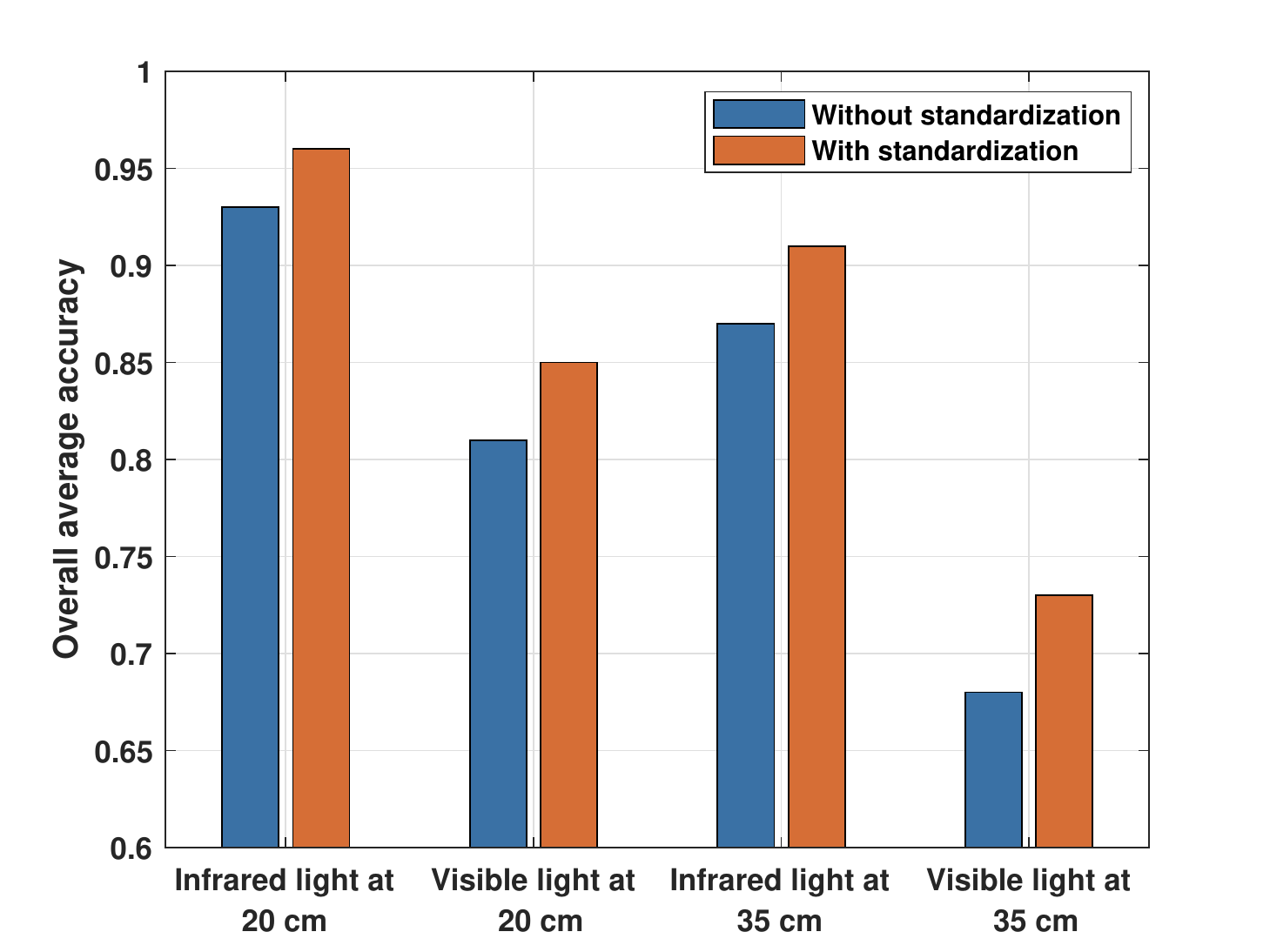}
\centering
\caption{Effect of standardization step on the average accuracy for both infrared and visible light source at 20~cm and 35~cm (ambient light is on).}
\label{fig:Accurcy_Standization_Effect}
\vspace{-2mm}
\end{figure}

However, the standardization discussed in Section \ref{sec:System_Design}. C was meant  to reduce the impact of absolute light intensity on performance.  The above result was therefore somewhat unexpected and compelled us to compare the recognition accuracy with and without standardization. As shown in Fig.~\ref{fig:Accurcy_Standization_Effect}, the standardization process  does not make the system agnostic to the impact of different sources and distances, but it does systematically increase the recognition accuracy.  
 
\begin{table}[t]
\centering
\begin{tabular}{@{}clcccccccc@{}}
\toprule
\multicolumn{1}{l}{}                                                                                 & \multicolumn{9}{c}{Estimating Gesture}                                                                                                                                                                                                           \\ \midrule
\multicolumn{1}{|c|}{\multirow{9}{*}{\begin{tabular}[c]{@{}c@{}}\rotatebox{90}{Performing Gesture\hspace{1cm}}\end{tabular}}} & \multicolumn{1}{l|}{}    & \multicolumn{1}{l|}{(a)} & \multicolumn{1}{l|}{(b)} & \multicolumn{1}{l|}{(c)} & \multicolumn{1}{l|}{(d)} & \multicolumn{1}{l|}{(e)} & \multicolumn{1}{l|}{(f)} & \multicolumn{1}{l|}{(g)} & \multicolumn{1}{l|}{(h)} \\ \cmidrule(l){2-10} 
\multicolumn{1}{|c|}{}                                                                               & \multicolumn{1}{l|}{(a)} &   \textbf{0.94}&0&0&0&0&0&0&0.06               \\ \cmidrule(lr){2-2}
\multicolumn{1}{|c|}{}                                                                               & \multicolumn{1}{l|}{(b)} &    0& \textbf{0.97}&0&0&0&0.03&0&0                     \\ \cmidrule(lr){2-2}
\multicolumn{1}{|c|}{}                                                                               & \multicolumn{1}{l|}{(c)} &   0 &0& \textbf{0.99}&0&0&0&0.01&0                        \\ \cmidrule(lr){2-2}
\multicolumn{1}{|c|}{}                                                                               & \multicolumn{1}{l|}{(d)} &    0&0.02&0& \textbf{0.98}&0&0&0&0                 \\ \cmidrule(lr){2-2}
\multicolumn{1}{|c|}{}                                                                               & \multicolumn{1}{l|}{(e)} &    0 &0&0&0& \textbf{0.98}&0&0.02&0           \\ \cmidrule(lr){2-2}
\multicolumn{1}{|c|}{}                                                                               & \multicolumn{1}{l|}{(f)} &    0.03 &0.02&0&0&0& \textbf{0.92}&0&0.03                 \\ \cmidrule(lr){2-2}
\multicolumn{1}{|c|}{}                                                                               & \multicolumn{1}{l|}{(g)} &    0&0&0.03&0&0&0& \textbf{0.97}&0                    \\ \cmidrule(lr){2-2}
\multicolumn{1}{|c|}{}                                                                               & \multicolumn{1}{l|}{(h)} &    0 &0&0&0&0&0&0& \textbf{1.00}      \\ \bottomrule
\end{tabular}
 \caption{The confusion matrix of infrared wave sensing around 20 cm in the dark (ambient light is off).}  
 \label{infdark}
\end{table}
\begin{table}[t]
\centering
\begin{tabular}{@{}clcccccccc@{}}
\toprule
\multicolumn{1}{l}{}                                                                                 & \multicolumn{9}{c}{Estimating Gesture}                                                                                                                                                                                                           \\ \midrule
\multicolumn{1}{|c|}{\multirow{9}{*}{\begin{tabular}[c]{@{}c@{}}\rotatebox{90}{Performing Gesture\hspace{1cm}}\end{tabular}}} & \multicolumn{1}{l|}{}    & \multicolumn{1}{l|}{(a)} & \multicolumn{1}{l|}{(b)} & \multicolumn{1}{l|}{(c)} & \multicolumn{1}{l|}{(d)} & \multicolumn{1}{l|}{(e)} & \multicolumn{1}{l|}{(f)} & \multicolumn{1}{l|}{(g)} & \multicolumn{1}{l|}{(h)} \\ \cmidrule(l){2-10} 
\multicolumn{1}{|c|}{}                                                                               & \multicolumn{1}{l|}{(a)} &  \textbf{0.85}  &       0      &   0   & 0.03  &  0 &   0.1 &   0.02  &  0              \\ \cmidrule(lr){2-2}
\multicolumn{1}{|c|}{}                                                                               & \multicolumn{1}{l|}{(b)} &     0 &   \textbf{0.90}   &  0    &    0.1  &  0   &     0  &  0   & 0                   \\ \cmidrule(lr){2-2}
\multicolumn{1}{|c|}{}                                                                               & \multicolumn{1}{l|}{(c)} &   0 &   0 &   \textbf{0.95}   &    0   & 0&     0     &    0.05  &  0                       \\ \cmidrule(lr){2-2}
\multicolumn{1}{|c|}{}                                                                               & \multicolumn{1}{l|}{(d)} &   0  &       0.03   & 0.06  &  \textbf{0.88} &  0  & 0.011  & 0 &  0                 \\ \cmidrule(lr){2-2}
\multicolumn{1}{|c|}{}                                                                               & \multicolumn{1}{l|}{(e)} &   0   & 0.03  &  0  & 0  &   \textbf{0.89}  &      0  &  0.08  &  0            \\ \cmidrule(lr){2-2}
\multicolumn{1}{|c|}{}                                                                               & \multicolumn{1}{l|}{(f)} &   0   &     0.04  &       0    &    0  &       0  &   \textbf{0.88} &  0.08 &  0.02                  \\ \cmidrule(lr){2-2}
\multicolumn{1}{|c|}{}                                                                               & \multicolumn{1}{l|}{(g)} &  0  &    0.01  &      0   & 0.01  &  0.03 &  0  &   \textbf{0.85} &  0.1                   \\ \cmidrule(lr){2-2}
\multicolumn{1}{|c|}{}                                                                               & \multicolumn{1}{l|}{(h)} &   0    &    0   & 0  &  0.07 &  0.01 &  0  &  0.06 &    \textbf{0.86}       \\ \bottomrule
\end{tabular}
 \caption{The confusion matrix of visible light-wave sensing  around 20 cm in the dark (ambient light is off).}  
 \label{visibledark}
  
\end{table}

\subsubsection{Accuracy with different ambient lighting}   

To determine the effect of ambient light on recognition accuracy, we compared the classifications result with ambient lights \emph{on} and \emph{off} using both infrared and visible light-wave sensing at $d=20$~cm. The ambient light is \emph{on} means the fluorescent lights in the ceiling and computer screen light are included in the environment light. The ambient light is \emph{off} means that there is no light in the room and the light sources are our infrared light or visible light source.

It is seen from the results in Tables~\ref{inf1} and \ref{infdark}, when using the infrared source, ambient lighting conditions have little, if any, significant impact.  This is consistent with the fact that the power of the reflected infrared light is much greater than the ambient contribution. 
However,  Tables~\ref{visible1} and \ref{visibledark} show that for visible light-wave sensing, recognition is slightly better when ambient lighting is \emph{off}.  This is attributed to the fact that the visible light source alongwith signal reflections are weaker. {The weaker reflected signals are impacted more by the environment noise compared with the stronger one.} 

As a related topic, we also show the performance of the denoising operation since this was meant to reduce the impact of noise level on recognition accuracy. 
As shown in Fig.~\ref{fig:Accurcy_denoising_Effect}, the accuracy systematically improves by including denoising, although the benefit is not uniform to all conditions. It appears to have the greatest positive impact as signal levels approach the environmental noise levels.

\section{Conclusion and Future Opportunities}
\label{sec:Conclusion}
In this paper, we presented a light-based hand gesture recognition system that utilized incoherent light reflection signals to accomplish  hand gesture recognition in a short range between  20~cm and 35~cm. The main innovation was the exploitation of ubiquitous light which is safe, low-cost, and easily generated and analyzed.  We have shown how we employ a series of signal processing steps and use machine learning such that this sensing modality can achieve high recognition accuracy for 8 gestures (in the case of infrared light sensing) in common ambient lighting conditions.  

In order to verify the limitations and capabilities of our system, more measurements and testing should be done on more subjects and in different lighting conditions.
The age, gender and skin complexion of subjects have to be taken into consideration.
Moreover, it was observed that recognition accuracy improved with transmitter power which suggested that improved performance can be achieved with greater SNR and/or dynamic range in the system. 
More photodetectors are needed to be applied to achieve the tracking of hand or body movement at large distances.
Future studies will be conducted to better quantify and model the operation of the system, to further verify this method's practicality and limitations, and to improve system performance.

\section*{Acknowledgment}
The authors would like to thank anonymous reviewers for their constructive comments which helped in improving this manuscript.

\ifCLASSOPTIONcaptionsoff
  \newpage
\fi

\bibliographystyle{IEEEtran}
\bibliography{my}

\begin{thebibliography}{10}
\providecommand{\url}[1]{#1}
\csname url@samestyle\endcsname
\providecommand{\newblock}{\relax}
\providecommand{\bibinfo}[2]{#2}
\providecommand{\BIBentrySTDinterwordspacing}{\spaceskip=0pt\relax}
\providecommand{\BIBentryALTinterwordstretchfactor}{4}
\providecommand{\BIBentryALTinterwordspacing}{\spaceskip=\fontdimen2\font plus
\BIBentryALTinterwordstretchfactor\fontdimen3\font minus
  \fontdimen4\font\relax}
\providecommand{\BIBforeignlanguage}[2]{{%
\expandafter\ifx\csname l@#1\endcsname\relax
\typeout{** WARNING: IEEEtran.bst: No hyphenation pattern has been}%
\typeout{** loaded for the language `#1'. Using the pattern for}%
\typeout{** the default language instead.}%
\else
\language=\csname l@#1\endcsname
\fi
#2}}
\providecommand{\BIBdecl}{\relax}
\BIBdecl

\bibitem{IOT1}
\BIBentryALTinterwordspacing
J.~Gubbi, R.~Buyya, S.~Marusic, and M.~Palaniswami, ``{Internet of Things
  (IoT): A vision, architectural elements, and future directions},''
  \emph{Future Generation Computer Systems}, vol.~29, no.~7, pp. 1645--1660,
  2013, including Special sections: Cyber-enabled Distributed Computing for
  Ubiquitous Cloud and Network Services \& Cloud Computing and Scientific
  Applications — Big Data, Scalable Analytics, and Beyond. [Online].
  Available:
  \url{http://www.sciencedirect.com/science/article/pii/S0167739X13000241}
\BIBentrySTDinterwordspacing

\bibitem{hcin1}
R.~Aigner, D.~Wigdor, H.~Benko, M.~Haller, D.~Lindlbauer, A.~Ion, S.~Zhao, and
  J.~Koh, ``{Understanding Mid-Air Hand Gestures: A Study of Human Preferences
  in Usage of Gesture Types for HCI},'' Nov 2012.

\bibitem{hcin2}
T.~{Seehapoch} and S.~{Wongthanavasu}, ``Speech emotion recognition using
  support vector machines,'' in \emph{2013 5th International Conference on
  Knowledge and Smart Technology (KST)}, 2013, pp. 86--91.

\bibitem{HCI4}
\BIBentryALTinterwordspacing
S.~S. Rautaray and A.~Agrawal, ``Vision based hand gesture recognition for
  human computer interaction: a survey,'' \emph{Artificial Intelligence
  Review}, vol.~43, no.~1, pp. 1--54, Jan 2015. [Online]. Available:
  \url{https://doi.org/10.1007/s10462-012-9356-9}
\BIBentrySTDinterwordspacing

\bibitem{wear1}
C.~Zhu and W.~Sheng, ``Wearable sensor-based hand gesture and daily activity
  recognition for robot-assisted living,'' \emph{IEEE Transactions on Systems,
  Man, and Cybernetics - Part A: Systems and Humans}, vol.~41, no.~3, pp.
  569--573, May 2011.

\bibitem{wear2}
A.~Nelson, J.~Schmandt, P.~Shyamkumar, W.~Wilkins, D.~Lachut, N.~Banerjee,
  S.~Rollins, J.~Parkerson, and V.~Varadan, ``Wearable multi-sensor gesture
  recognition for paralysis patients,'' in \emph{2013 IEEE SENSORS}, Nov 2013,
  pp. 1--4.

\bibitem{wear3}
Z.~Lv, ``Wearable smartphone: Wearable hybrid framework for hand and foot
  gesture interaction on smartphone,'' in \emph{2013 IEEE International
  Conference on Computer Vision Workshops}, Dec 2013, pp. 436--443.

\bibitem{wear4}
M.~Caputo, K.~Denker, B.~Dums, G.~Umlauf, H.~Konstanz, and G.~, ``{3D Hand
  Gesture Recognition Based on Sensor Fusion of Commodity Hardware},'' vol.
  2012, Jan 2012.

\bibitem{light7}
\BIBentryALTinterwordspacing
R.~H. Venkatnarayan and M.~Shahzad, ``Gesture recognition using ambient
  light,'' \emph{Proc. ACM Interact. Mob. Wearable Ubiquitous Technol.},
  vol.~2, no.~1, pp. 40:1--40:28, Mar. 2018. [Online]. Available:
  \url{http://doi.acm.org/10.1145/3191772}
\BIBentrySTDinterwordspacing

\bibitem{RF1}
\BIBentryALTinterwordspacing
F.~Adib, Z.~Kabelac, D.~Katabi, and R.~C. Miller, ``3d tracking via body radio
  reflections,'' in \emph{11th {USENIX} Symposium on Networked Systems Design
  and Implementation ({NSDI} 14)}.\hskip 1em plus 0.5em minus 0.4em\relax
  Seattle, WA: {USENIX} Association, 2014, pp. 317--329. [Online]. Available:
  \url{https://www.usenix.org/conference/nsdi14/technical-sessions/presentation/adib}
\BIBentrySTDinterwordspacing

\bibitem{RF3}
\BIBentryALTinterwordspacing
S.~Sen, J.~Lee, K.-H. Kim, and P.~Congdon, ``{Avoiding Multipath to Revive
  Inbuilding WiFi Localization},'' in \emph{Proceeding of the 11th Annual
  International Conference on Mobile Systems, Applications, and Services}, ser.
  MobiSys '13.\hskip 1em plus 0.5em minus 0.4em\relax New York, NY, USA: ACM,
  2013, pp. 249--262. [Online]. Available:
  \url{http://doi.acm.org/10.1145/2462456.2464463}
\BIBentrySTDinterwordspacing

\bibitem{RF4}
\BIBentryALTinterwordspacing
J.~Lien, N.~Gillian, M.~E. Karagozler, P.~Amihood, C.~Schwesig, E.~Olson,
  H.~Raja, and I.~Poupyrev, ``Soli: Ubiquitous gesture sensing with millimeter
  wave radar,'' \emph{ACM Trans. Graph.}, vol.~35, no.~4, pp. 142:1--142:19,
  Jul. 2016. [Online]. Available:
  \url{http://doi.acm.org/10.1145/2897824.2925953}
\BIBentrySTDinterwordspacing

\bibitem{emirf1}
Z.~Chi, Y.~Yao, T.~Xie, X.~Liu, Z.~Huang, W.~Wang, and T.~Zhu, ``{EAR:
  Exploiting uncontrollable ambient RF signals in heterogeneous networks for
  gesture recognition},'' in \emph{{SenSys 2018 - Proceedings of the 16th
  Conference on Embedded Networked Sensor Systems}}, Nov 2018, pp. 237--249.

\bibitem{emirf2}
Z.~Tian, X.~Yang, and M.~Zhou, ``{WiCatch: A Wi-Fi Based Hand Gesture
  Recognition System},'' \emph{IEEE Access}, vol.~6, pp. 16\,911--16\,923, Mar
  2018.

\bibitem{image1}
\BIBentryALTinterwordspacing
L.~Chen, H.~Wei, and J.~Ferryman, ``A survey of human motion analysis using
  depth imagery,'' \emph{Pattern Recognition Letters}, vol.~34, no.~15, pp.
  1995 -- 2006, 2013, smart Approaches for Human Action Recognition. [Online].
  Available:
  \url{http://www.sciencedirect.com/science/article/pii/S0167865513000500}
\BIBentrySTDinterwordspacing

\bibitem{image2}
\BIBentryALTinterwordspacing
M.~A.~R. Ahad, J.~K. Tan, H.~Kim, and S.~Ishikawa, ``Motion history image: Its
  variants and applications,'' \emph{Mach. Vision Appl.}, vol.~23, no.~2, p.
  255–281, Mar. 2012. [Online]. Available:
  \url{https://doi.org/10.1007/s00138-010-0298-4}
\BIBentrySTDinterwordspacing

\bibitem{deep1}
M.~{Asadi-Aghbolaghi}, A.~{Clapés}, M.~{Bellantonio}, H.~J. {Escalante},
  V.~{Ponce-López}, X.~{Baró}, I.~{Guyon}, S.~{Kasaei}, and S.~{Escalera},
  ``{A Survey on Deep Learning Based Approaches for Action and Gesture
  Recognition in Image Sequences},'' in \emph{{2017 12th IEEE International
  Conference on Automatic Face Gesture Recognition (FG 2017)}}, Washington, DC,
  USA, 2017, pp. 476--483.

\bibitem{deep2}
O.~Oyedotun and A.~Khashman, ``Deep learning in vision-based static hand
  gesture recognition,'' \emph{Neural Computing and Applications}, vol.~28, pp.
  3941--3951, Apr 2016.

\bibitem{camera5}
S.~Oprisescu, C.~Rasche, and B.~Su, ``Automatic static hand gesture recognition
  using tof cameras,'' in \emph{2012 Proceedings of the 20th European Signal
  Processing Conference (EUSIPCO)}, Aug 2012, pp. 2748--2751.

\bibitem{camera6}
T.~Plotz, C.~Chen, N.~Y. Hammerla, and G.~D. Abowd, ``Automatic synchronization
  of wearable sensors and video-cameras for ground truth annotation -- a
  practical approach,'' in \emph{2012 16th International Symposium on Wearable
  Computers}, Jun 2012, pp. 100--103.

\bibitem{soundnew1}
Y.~{Qifan}, T.~{Hao}, Z.~{Xuebing}, L.~{Yin}, and Z.~{Sanfeng}, ``Dolphin:
  Ultrasonic-based gesture recognition on smartphone platform,'' in \emph{2014
  IEEE 17th International Conference on Computational Science and Engineering},
  2014, pp. 1461--1468.

\bibitem{soundnew2}
\BIBentryALTinterwordspacing
A.~Mujibiya, X.~Cao, D.~S. Tan, D.~Morris, S.~N. Patel, and J.~Rekimoto, ``The
  sound of touch: On-body touch and gesture sensing based on transdermal
  ultrasound propagation,'' in \emph{Proceedings of the 2013 ACM International
  Conference on Interactive Tabletops and Surfaces}, ser. ITS ’13.\hskip 1em
  plus 0.5em minus 0.4em\relax New York, NY, USA: Association for Computing
  Machinery, 2013, p. 189–198. [Online]. Available:
  \url{https://doi.org/10.1145/2512349.2512821}
\BIBentrySTDinterwordspacing

\bibitem{soundnew}
\BIBentryALTinterwordspacing
H.~Watanabe, T.~Terada, and M.~Tsukamoto, ``{Ultrasound-Based Movement Sensing,
  Gesture-, and Context-Recognition},'' in \emph{Proceedings of the 2013
  International Symposium on Wearable Computers}, ser. ISWC ’13.\hskip 1em
  plus 0.5em minus 0.4em\relax New York, NY, USA: Association for Computing
  Machinery, 2013, p. 57–64. [Online]. Available:
  \url{https://doi.org/10.1145/2493988.2494335}
\BIBentrySTDinterwordspacing

\bibitem{light6}
\BIBentryALTinterwordspacing
C.~Zhang, J.~Tabor, J.~Zhang, and X.~Zhang, ``Extending mobile interaction
  through near-field visible light sensing,'' in \emph{Proceedings of the 21st
  Annual International Conference on Mobile Computing and Networking}, ser.
  MobiCom '15.\hskip 1em plus 0.5em minus 0.4em\relax New York, NY, USA: ACM,
  2015, pp. 345--357. [Online]. Available:
  \url{http://doi.acm.org/10.1145/2789168.2790115}
\BIBentrySTDinterwordspacing

\bibitem{light2}
Y.~Yang, J.~Hao, J.~Luo, and S.~J. Pan, ``Ceilingsee: Device-free occupancy
  inference through lighting infrastructure based led sensing,'' in \emph{2017
  IEEE International Conference on Pervasive Computing and Communications
  (PerCom)}, March 2017, pp. 247--256.

\bibitem{light16}
M.~Kaholokula, ``Reusing ambient light to recognize hand gestures,''
  \emph{Dartmouth College}, 2016.

\bibitem{2s}
\BIBentryALTinterwordspacing
T.~Li, X.~Xiong, Y.~Xie, G.~Hito, X.-D. Yang, and X.~Zhou, ``Reconstructing
  hand poses using visible light,'' \emph{Proceedings of the ACM on
  Interactive, Mobile, Wearable and Ubiquitous Technologies}, vol.~1, no.~3,
  Sep 2017. [Online]. Available: \url{https://doi.org/10.1145/3130937}
\BIBentrySTDinterwordspacing

\bibitem{light3}
T.~Hao, R.~Zhou, and G.~Xing, ``Cobra: color barcode streaming for smartphone
  systems,'' in \emph{MobiSys}, 2012.

\bibitem{light4}
H.~{Cheng}, A.~M. {Chen}, A.~{Razdan}, and E.~{Buller}, in \emph{2011 IEEE
  International Conference on Consumer Electronics (ICCE)}, Jan 2011, pp.
  149--150.

\bibitem{1s}
\BIBentryALTinterwordspacing
J.~Gong, Y.~Zhang, X.~Zhou, and X.-D. Yang, ``Pyro: Thumb-tip gesture
  recognition using pyroelectric infrared sensing,'' in \emph{Proceedings of
  the 30th Annual ACM Symposium on User Interface Software and Technology},
  ser. UIST ’17.\hskip 1em plus 0.5em minus 0.4em\relax New York, NY, USA:
  Association for Computing Machinery, 2017, p. 553–563. [Online]. Available:
  \url{https://doi.org/10.1145/3126594.3126615}
\BIBentrySTDinterwordspacing

\bibitem{3s}
\BIBentryALTinterwordspacing
Y.~Li, T.~Li, R.~A. Patel, X.-D. Yang, and X.~Zhou, ``Self-powered gesture
  recognition with ambient light,'' in \emph{Proceedings of the 31st Annual ACM
  Symposium on User Interface Software and Technology}, ser. UIST ’18.\hskip
  1em plus 0.5em minus 0.4em\relax New York, NY, USA: Association for Computing
  Machinery, 2018, p. 595–608. [Online]. Available:
  \url{https://doi.org/10.1145/3242587.3242635}
\BIBentrySTDinterwordspacing

\bibitem{nist}
C.~C. Cooksey and D.~W. Allen, ``Reflectance measurements of human skin from
  the ultraviolet to the shortwave infrared (250 nm to 2500 nm),'' in
  \emph{Defense, Security, and Sensing}, 2013.

\bibitem{transmitter}
\BIBentryALTinterwordspacing
{940nm IR lamp Board with Light Sensor (48 Black LED Illuminator Array)}.
  Amazon.com, Inc. Accessed on: 07-11-2020. [Online]. Available:
  \url{https://www.amazon.com/gp/product/B0785W2RQQ}
\BIBentrySTDinterwordspacing

\bibitem{vistran}
\BIBentryALTinterwordspacing
{Super Bright White 5mm LED (25 pack)}. Adafruit Industries. Accessed on:
  07-11-2020. [Online]. Available: \url{https://www.adafruit.com/product/754}
\BIBentrySTDinterwordspacing

\bibitem{detector}
\BIBentryALTinterwordspacing
{PDA100A}. Thorlabs, Inc. Accessed on: 07-11-2020. [Online]. Available:
  \url{https://www.thorlabs.com/thorproduct.cfm?partnumber=PDA100A}
\BIBentrySTDinterwordspacing

\bibitem{dwtn1}
S.~{Lahmiri} and M.~{Boukadoum}, ``Physiological signal denoising with
  variational mode decomposition and weighted reconstruction after dwt
  thresholding,'' in \emph{2015 IEEE International Symposium on Circuits and
  Systems (ISCAS)}, 2015, pp. 806--809.

\bibitem{ecgnew}
\BIBentryALTinterwordspacing
P.~Singh, G.~Pradhan, and S.~S., ``{Denoising of ECG signal by non-local
  estimation of approximation coefficients in DWT},'' \emph{Biocybernetics and
  Biomedical Engineering}, vol.~37, no.~3, pp. 599 -- 610, 2017. [Online].
  Available:
  \url{http://www.sciencedirect.com/science/article/pii/S0208521617301122}
\BIBentrySTDinterwordspacing

\bibitem{dwtnew1}
\BIBentryALTinterwordspacing
J.~P. Amezquita-Sanchez and H.~Adeli, ``A new music-empirical wavelet transform
  methodology for time–frequency analysis of noisy nonlinear and
  non-stationary signals,'' \emph{Digital Signal Processing}, vol.~45, pp. 55
  -- 68, 2015. [Online]. Available:
  \url{http://www.sciencedirect.com/science/article/pii/S1051200415001992}
\BIBentrySTDinterwordspacing

\bibitem{thold}
A.~{Kashaf}, N.~{Javaid}, Z.~A. {Khan}, and I.~A. {Khan}, ``Tsep:
  Threshold-sensitive stable election protocol for wsns,'' in \emph{2012 10th
  International Conference on Frontiers of Information Technology}, 2012, pp.
  164--168.

\bibitem{thhhh}
J.~Lord, M.~Rast, C.~Mckinlay, and P.~Mininni, ``{Wavelet decomposition of
  forced turbulence: Applicability of the iterative Donoho-Johnstone
  threshold},'' \emph{Physics of Fluids}, vol.~24, Feb 2012.

\bibitem{th3}
D.~{Valencia}, D.~{Orejuela}, J.~{Salazar}, and J.~{Valencia}, ``Comparison
  analysis between rigrsure, sqtwolog, heursure and minimaxi techniques using
  hard and soft thresholding methods,'' in \emph{2016 XXI Symposium on Signal
  Processing, Images and Artificial Vision}, Aug 2016, pp. 1--5.

\bibitem{dtw_}
F.~{Zhou} and F.~{De la Torre}, ``Generalized time warping for multi-modal
  alignment of human motion,'' in \emph{2012 IEEE Conference on Computer Vision
  and Pattern Recognition}, 2012, pp. 1282--1289.

\bibitem{hand1}
W.~T. Freeman and M.~Roth, ``Orientation histograms for hand gesture
  recognition,'' Mitsubishi Electric Research Labs., 201, Tech. Rep., 213.

\bibitem{knn11}
Q.~{Chen}, D.~{Li}, and C.~{Tang}, ``Knn matting,'' \emph{IEEE Transactions on
  Pattern Analysis and Machine Intelligence}, vol.~35, no.~9, pp. 2175--2188,
  2013.

\bibitem{spectro}
\BIBentryALTinterwordspacing
{USB Series UV-NIR Spectrometers}. Ocean Insight. Accessed on: 07-11-2020.
  [Online]. Available:
  \url{https://www.oceaninsight.com/products/spectrometers/usb-series/usb-uv-nir/}
\BIBentrySTDinterwordspacing

\bibitem{pda}
\BIBentryALTinterwordspacing
{PDA100A(-EC) Si Switchable Gain Detector User Guide}. Thorlabs, Inc. Accessed
  on: 07-11-2020. [Online]. Available:
  \url{https://www.thorlabs.com/drawings/4db5f4acdbba82df-2EE78459-B895-BFEB-3C3F1B3F7149ECEF/PDA100A-Manual.pdf}
\BIBentrySTDinterwordspacing

\end{thebibliography}

%%%%%%%%%%%%%%%%%%%%%%%%%%%%%%%%%%%%%%%%%%%%%%
%%%%%%%%%%%% SECTION %%%%%%%%%%%%%%%%%%%%%%%%%
%%%%%%%%%%%%%%%%%%%%%%%%%%%%%%%%%%%%%%%%%%%%%% 
%Where the bibliography will be printed

\vskip 0pt plus -1fil
%\vspace*{-8mm}
\begin{IEEEbiography}[{ \includegraphics[width=1in,height=1.2in,clip]{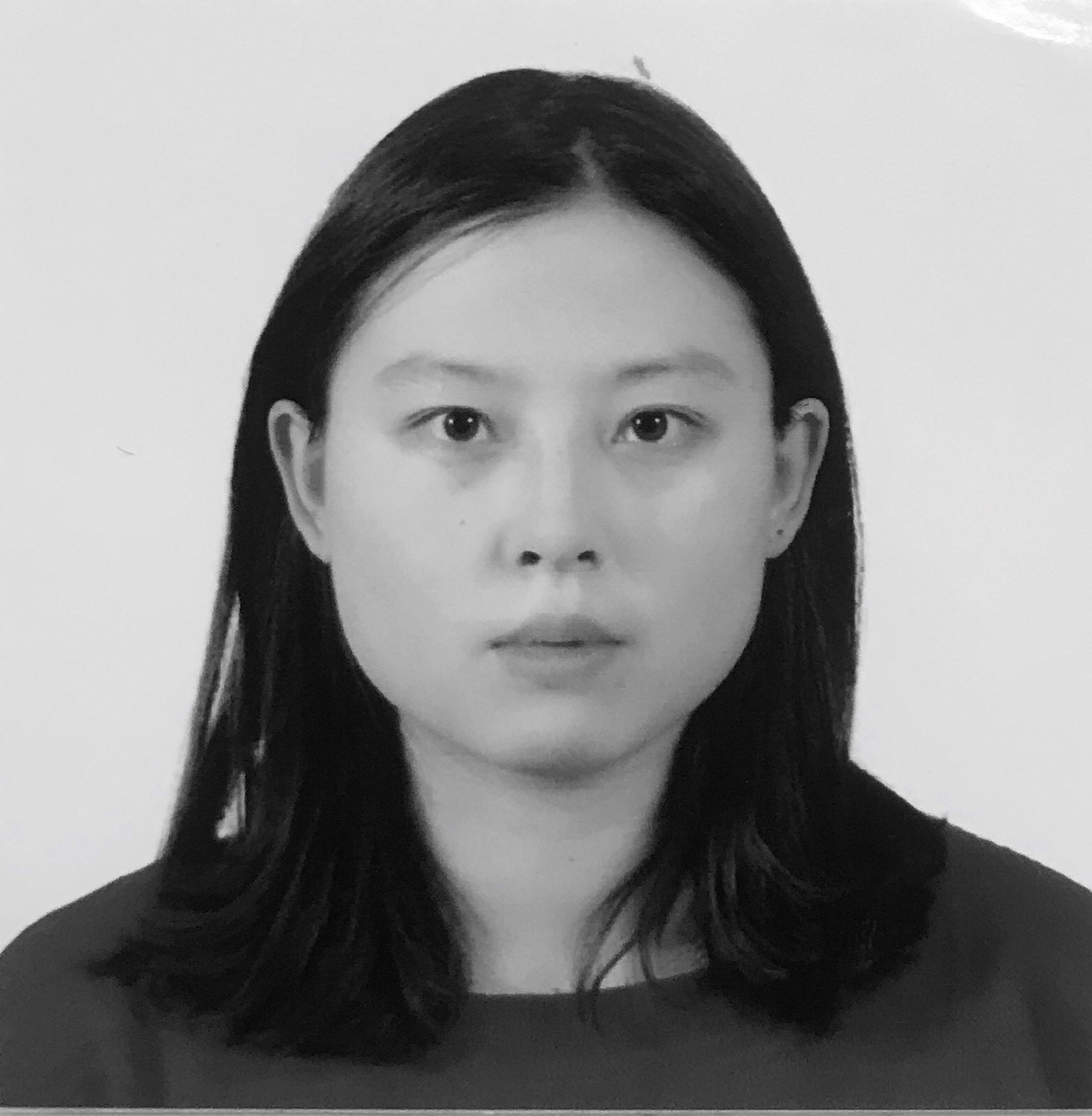}}]{Li Yu} 
received the B.Sc. degree in guidance and control technology from Harbin Engineering University, Harbin, China, in 2013, and the M.Sc. degree in control science and engineering from Beijing Institute of Technology, Beijing, China, in 2016, and the M.Sc. degree in electrical and computer engineering from Oklahoma State University, OK, USA, in 2019. Her research interests include the gesture recognition, visible light sensing and machine learning.
% \vspace*{-10mm}
\end{IEEEbiography}
% \vskip 0pt plus -1fil

\begin{IEEEbiography}[{ \includegraphics[width=1in,height=1.2in,clip]{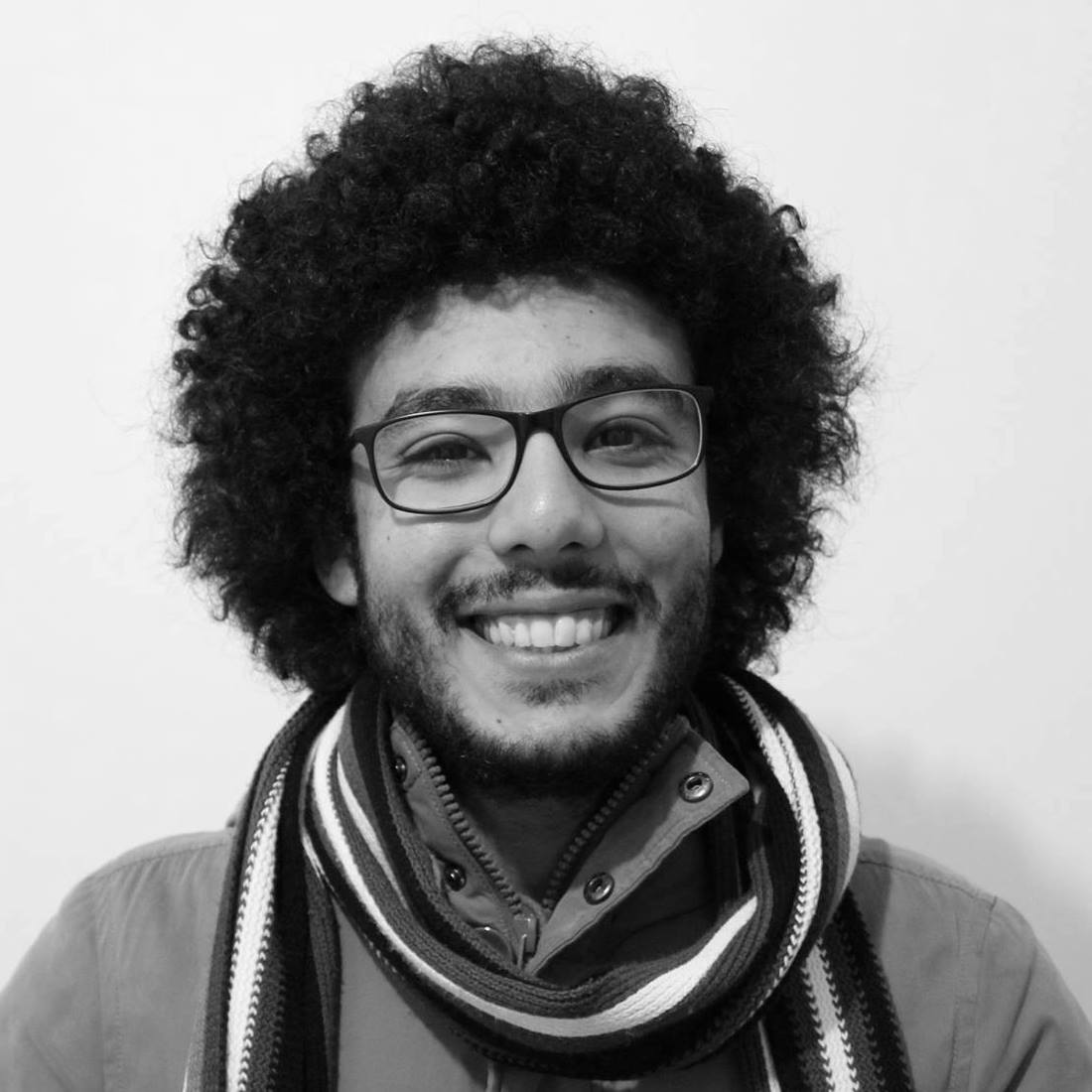}} ]{Hisham Abuella} received the B.Sc. degree in Communications and Electronics Engineering from Ain Shams University, Cairo, Egypt, in 2013. He worked as a Digital System Design Engineer in Varkon Semiconductor (Wasiela) Company, Cairo, Egypt. In Fall 2014, he joined Istanbul Sehir University as a Research Assistant for his M.Sc. degree in Electronics and Computer Engineering from  Istanbul Sehir University, Turkey. Lastly, he joined Oklahoma State University as a Graduate Research Assistant to pursue his Ph.D. study at the School of Electrical and Computer Engineering in Spring 2017. He was an engineering intern in Qualcomm Inc., San Diego, CA during the summer of 2019. His current research interests include light sensing and communication, wireless communication systems design using SDRs, visible light Sensing applications, and machine learning and DSP algorithms for wireless communication systems.
% \vspace*{-10mm}
\end{IEEEbiography}
% \vskip 0pt plus -1fil

\begin{IEEEbiography}[{ \includegraphics[width=1in,height=1.2in,clip]{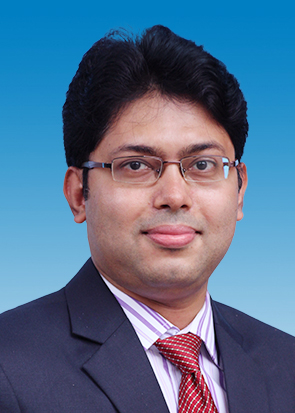}} ]{Md Zobaer Islam} received the B.Sc. degree in Electrical and Electronic Engineering from Bangladesh University of Engineering and Technology, in 2012. He joined Oklahoma State University as a Graduate Teaching Assistant to pursue his Ph.D. study at the School of Electrical and Computer Engineering in Spring 2020. He has industry experience of 4 years at Bangladesh Telecommunications Company Ltd. in telecommunication and information technology (IT) area and 3 years at Samsung R\&D Institute Bangladesh in software area. His current research interests include wireless light-wave sensing system design and cognitive radio systems.
\vspace*{-10mm}
\end{IEEEbiography}
%\vskip 0pt plus -1fil

\begin{IEEEbiography}[{ \includegraphics[width=1in,height=1.2in,clip]{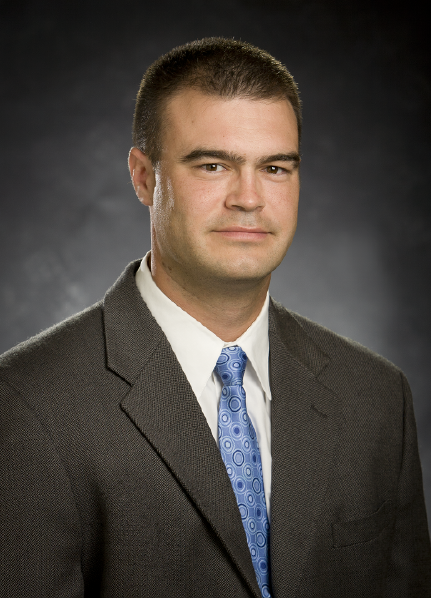}}]{John F. O'Hara (SM'19)} received his BSEE degree from the University of Michigan in 1998 and his Ph.D. (electrical engineering) from Oklahoma State University in 2003.  He was a Director of Central Intelligence Postdoctoral Fellow at Los Alamos National Laboratory (LANL) until 2006.  From 2006-2011 he was with the Center for Integrated Nanotechnologies (LANL) and worked on numerous metamaterial projects involving dynamic control over chirality, resonance frequency, polarization, and modulation of terahertz waves.  In 2011, he founded a consulting/research company, Wavetech, LLC.  In 2017, he joined Oklahoma State University as an Assistant Professor, where he now studies metamaterials, terahertz communications, and photonic sensing technologies.   He has around 100 publications in journals and conference proceedings.
% \vspace*{-10mm}
\end{IEEEbiography}
% \vskip 0pt plus -1fil

\begin{IEEEbiography}[{ \includegraphics[width=1in,height=1.2in,clip]{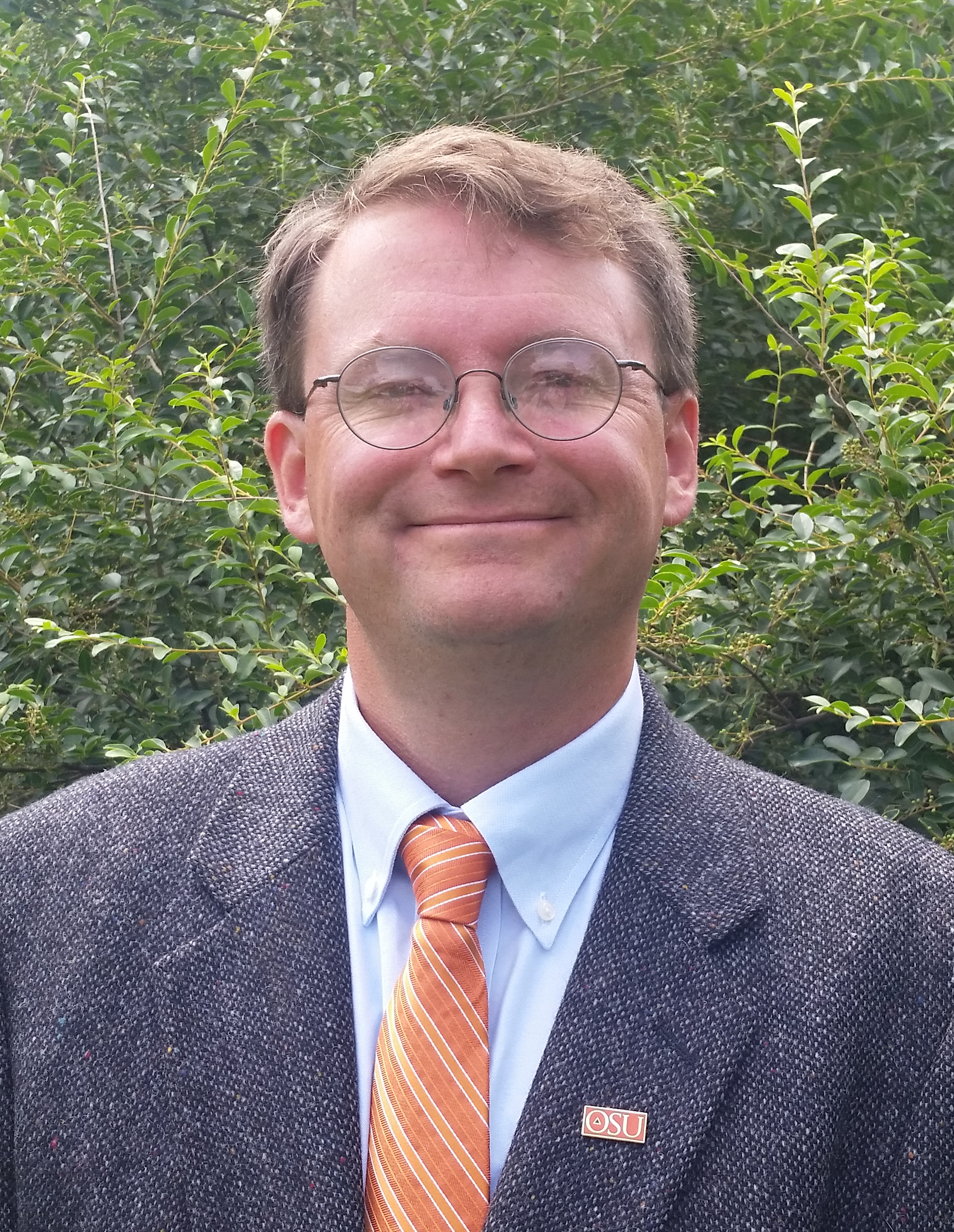}}]{Christopher Crick} received his Ph.D. from Yale University in 2009. He is currently Associate Professor of Computer Science at Oklahoma State University and director of the Cognitive Robotics Laboratory. He is the author of nearly fifty articles in the subject areas of robotics, cognitive science, artificial intelligence and machine learning.  His work addresses the many challenges of humans and robots working together in teams.
% \vspace*{-10mm}
\end{IEEEbiography}
% \vskip 0pt plus -1fil

\begin{IEEEbiography}[{
\includegraphics[width=1in,height=1.2in,clip]{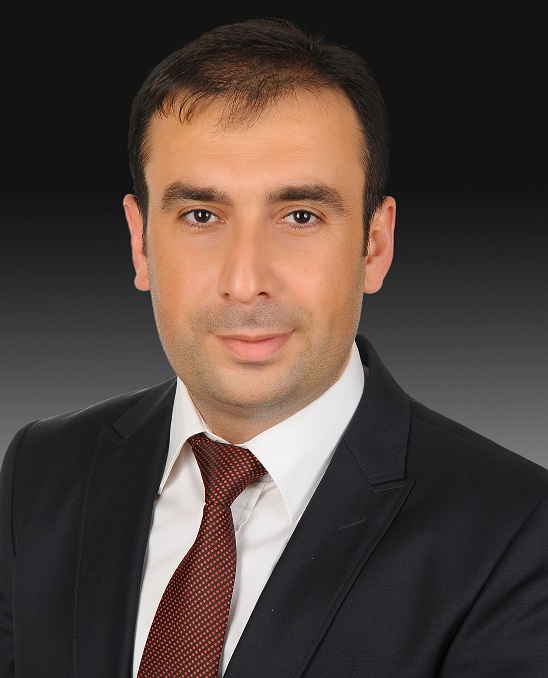}}]{Sabit Ekin (M'12)}  received the B.Sc. degree in electrical and electronics engineering from Eski\c sehir Osmangazi University, Turkey, in 2006, the M.Sc. degree in electrical engineering from New Mexico Tech, Socorro, NM, USA, in 2008, and the Ph.D. degree in electrical and computer engineering from Texas A\&M University, College Station, TX, USA, in 2012. In summer 2012, he was with the Femtocell Interference Management Team in the Corporate Research and Development, New Jersey Research Center, Qualcomm Inc. He joined the School of Electrical and Computer Engineering, Oklahoma State University, Stillwater, OK, USA, as an Assistant Professor, in 2016. He has four years of industrial experience from Qualcomm Inc., as a Senior Modem Systems Engineer with the Department of Qualcomm Mobile Computing. At Qualcomm Inc., he has received numerous Qualstar awards for his achievements/contributions on cellular modem receiver design. His research interests include the design and performance analysis of wireless  systems in both theoretical and practical point of views, visible light sensing, communications and applications, non-contact health monitoring, and Internet of Things applications.
\end{IEEEbiography}

\end{document}